\documentclass[preprint2]{aastex}

\newcommand{\br}{\mbox{$B\!-\!R$}}     
\newcommand{\ri}{\mbox{$R\!-\!i'$}}    
\newcommand{\vi}{\mbox{$V\!-\!i'$}}    
\newcommand{\iz}{\mbox{$i'\!-\!z'$}}   
\newcommand{\gr}{\mbox{$G\!-\!\mathfrak{R}$}}   
\newcommand{\rrii}{\mbox{$\mathfrak{R}\!-\!I$}} 
\newcommand{\ric}{\mbox{$R\!-\!I_c$}}  
\newcommand{\vic}{\mbox{$V\!-\!I_c$}}  
\newcommand{\icz}{\mbox{$I_c\!-\!z'$}} 
\newcommand{\izr}{\mbox{$i'\!-\!z_\mathrm{R}$}} 
\newcommand{\zbzr}{\mbox{$z_\mathrm{B}\!-\!z_\mathrm{R}$}} 
\newcommand{\iizz}{\mbox{$i\!-\!z$}}   
\newcommand{\Vz}{\mbox{$V\!-\!z$}}     
\newcommand{\ug}{\mbox{$U_n\!-\!G$}}   

\newcommand{\reffig}[1]{Figure~\ref{fig:#1}}
\newcommand{\reftbl}[1]{Table~\ref{tbl:#1}}

\begin{document}

\title{%
Luminosity Functions of Lyman-Break Galaxies at $z\sim 4$ and 5\\
in the Subaru Deep Field
\altaffilmark{1}
}

\author{%
Makiko Yoshida~\altaffilmark{2},
Kazuhiro Shimasaku~\altaffilmark{2,3},
Nobunari Kashikawa~\altaffilmark{4,5},
Masami Ouchi~\altaffilmark{6,7},
Sadanori Okamura~\altaffilmark{2,3},
Masaru Ajiki~\altaffilmark{8},
Masayuki Akiyama~\altaffilmark{9},
Hiroyasu Ando~\altaffilmark{4},
Kentaro Aoki~\altaffilmark{9},
Mamoru Doi~\altaffilmark{10},
Hisanori Furusawa~\altaffilmark{9},
Tomoki Hayashino~\altaffilmark{11},
Fumihide Iwamuro~\altaffilmark{12},
Masanori Iye~\altaffilmark{4,5},
Hiroshi Karoji~\altaffilmark{12},
Naoto Kobayashi~\altaffilmark{10},
Keiichi Kodaira~\altaffilmark{13},
Tadayuki Kodama~\altaffilmark{4},
Yutaka Komiyama~\altaffilmark{4},
Matthew A. Malkan~\altaffilmark{14},
Yuichi Matsuda~\altaffilmark{12},
Satoshi Miyazaki~\altaffilmark{9},
Yoshihiko Mizumoto~\altaffilmark{4},
Tomoki Morokuma~\altaffilmark{10},
Kentaro Motohara~\altaffilmark{10},
Takashi Murayama~\altaffilmark{8},
Tohru Nagao~\altaffilmark{4,15},
Kyoji Nariai~\altaffilmark{16},
Kouji Ohta~\altaffilmark{12},
Toshiyuki Sasaki~\altaffilmark{9},
Yasunori Sato~\altaffilmark{4},
Kazuhiro Sekiguchi~\altaffilmark{9},
Yasuhiro Shioya~\altaffilmark{8},
Hajime Tamura~\altaffilmark{11},
Yoshiaki Taniguchi~\altaffilmark{8},
Masayuki Umemura~\altaffilmark{17},
Toru Yamada~\altaffilmark{4},
and
Naoki Yasuda~\altaffilmark{18}
}

\email{myoshida@astron.s.u-tokyo.ac.jp}

\altaffiltext{1}{%
Based on data collected at the Subaru Telescope,
which is operated by the National Astronomical Observatory of Japan.}
\altaffiltext{2}{%
Department of Astronomy, Graduate School of Science,
The University of Tokyo, Tokyo 113-0033, Japan}
\altaffiltext{3}{%
Research Center for the Early Universe, Graduate School of Science,
The University of Tokyo, Tokyo 113-0033, Japan}
\altaffiltext{4}{%
Optical and Infrared Astronomy Division,
National Astronomical Observatory of Japan,
Mitaka, Tokyo 181-8588, Japan}
\altaffiltext{5}{%
Department of Astronomy, School of Science,
Graduate University for Advanced Studies,
Mitaka, Tokyo 181-8588, Japan}
\altaffiltext{6}{%
Space Telescope Science Institute,
3700, San Martin Drive, Baltimore, MD 21218, USA}
\altaffiltext{7}{%
Hubble Fellow}
\altaffiltext{8}{%
Astronomical Institute, Graduate School of Science,
Tohoku University, Aramaki, Aoba, Sendai 980-8578, Japan}
\altaffiltext{9}{%
Subaru Telescope, National Astronomical Observatory of Japan, 
650 N. A'ohoku Place, Hilo, HI 96720, USA}
\altaffiltext{10}{%
Institute of Astronomy, Graduate School of Science,
The University of Tokyo, Mitaka, Tokyo 181-8588, Japan}
\altaffiltext{11}{%
Research Center for Neutrino Science, Graduate School of Science,
Tohoku University, Aramaki, Aoba, Sendai 980-8578, Japan}
\altaffiltext{12}{%
Department of Astronomy, Graduate School of Science,
Kyoto University, Kyoto 606-8502, Japan}
\altaffiltext{13}{%
Graduate University for Advanced Studies,
Shonan Village, Hayama, Kanagawa 240-0193, Japan}
\altaffiltext{14}{%
Department of Physics and Astronomy, University of California,
Los Angeles, CA 90095, USA}
\altaffiltext{15}{%
INAF -- Osservatorio Astrofisico di Arcetri,
Largo Enrico Fermi 5, 50125 Firenze, Italy}
\altaffiltext{16}{%
Department of Physics, Meisei University,
2-1-1 Hodokubo, Hino, Tokyo 191-8506, Japan}
\altaffiltext{17}{%
Center for Computational Physics, University of Tsukuba,
1-1-1 Tennodai, Tsukuba 305-8571, Japan}
\altaffiltext{18}{%
Institute for Cosmic Ray Research, The University of Tokyo,
Kashiwa, Chiba, 277-8582, Japan}

\begin{abstract}
We investigate the luminosity functions of Lyman-break galaxies (LBG)
at $z\sim 4$ and 5
based on the optical imaging data obtained in the Subaru
Deep Field (SDF) Project,
a program conducted by Subaru Observatory
to carry out a deep
and wide survey of distant galaxies.
Three samples of LBGs
in a contiguous 875 arcmin$^2$ area
are constructed.
One consists of LBGs at $z\sim 4$ down to $i'=26.85$
selected with the \br\ vs \ri\ 
diagram ($BRi'$-LBGs).
The other two consist of LBGs at $z\sim 5$
down to $z'=26.05$
selected with two kinds of
two-color diagrams: \vi\ vs \iz \  ($Vi'z'$-LBGs) and \ri\ vs
\iz\ ($Ri'z'$-LBGs).
The number detected is $3,808$ for $BRi'$-LBGs, 539 for
$Vi'z'$-LBGs, and 240 for $Ri'z'$-LBGs.
The adopted selection criteria are proved to be fairly reliable
by the spectroscopic observation of 63 LBG candidates,
among which only 2 are found to be foreground objects.
We estimate the fraction of contamination and the completeness
for these three samples by Monte Carlo simulations, and derive the
luminosity functions of the LBGs at rest-frame ultraviolet
wavelengths down to $M_\mathrm{UV}=-19.2$ at $z\sim 4$ and
$M_\mathrm{UV}=-20.3$ at $z\sim 5$.
We find clear evolution of the luminosity function over the
redshift range of $0\le z\le 6$,
which is accounted for by a sole change in
the characteristic magnitude, $M^\ast$.
The cosmic star formation rate (SFR) density at $z\sim 4$ and
$z\sim 5$ is measured from
the luminosity functions.
The measurements of the integrated SFR density at these redshifts are
largely improved, since the luminosity functions are derived down to
very faint magnitudes.
We examine the evolution of the cosmic SFR density
and its luminosity dependence
over $0\le z\lesssim 6$.
The SFR density contributed from brighter
galaxies is found to change more drastically with cosmic time.
The contribution from
galaxies brighter than $M^\ast_\mathrm{z=3}-0.5$
has a sharp peak around
$z=3$ -- 4,
while that from
galaxies fainter than $M^\ast_\mathrm{z=3}-0.5$
evolves relatively mildly
with a broad peak at earlier epoch.
Combining the observed SFR density with
the standard Cold Dark Matter model, we compute the cosmic SFR
per unit baryon mass in dark haloes, i.e., the specific SFR.
The specific SFR is found to scale with redshift as $(1+z)^3$
up to $z\sim 4$, implying that
the efficiency of star formation is on average
higher at higher redshift in proportion to the cooling rate
within dark haloes,
while this is not simply the case at $z\gtrsim 4$.
\end{abstract}

\keywords{%
cosmology: observations ---
galaxies: evolution ---
galaxies: high-redshift ---
galaxies: luminosity function, mass function}

\section{INTRODUCTION} \label{sec:intro}

When and how galaxies formed is one of the primary questions
in astronomy today.
Observations of young galaxies at high redshifts
are a straightforward approach to this problem.
Over the past decade, it has become feasible to undertake large
surveys of galaxies at high redshifts.
What made this possible is the progress in observing technology
including telescopes and detectors, and also sophistication of
selection methods to locate high-redshift galaxies.
Probably the most efficient method is the Lyman-break technique.
It is a simple photometric technique based on the continuum features
in rest-frame ultraviolet spectra redshifted into optical bandpasses,
and requires only optical imaging in a few bandpasses.
Pioneered by \citet{guhathakurta1990},
this method has been successfully
used to find many young, star-forming galaxies beyond $z\sim 2$
\citep[e.g.,][]{steidel1992, steidel1999,
lehnert2003, iwata2003, ouchi2004, dickinson2004,
sawicki2006}.
\citet{steidel2003} made spectroscopic observation for about 1000
photometrically selected $z\sim 3$ galaxies and verified the
usefulness of this method.
High-redshift galaxies selected by this method are called Lyman-break
galaxies (LBGs).

Since their first discovery in the 1990s, various properties of
LBGs have been extensively studied.
One of the most fundamental measurements is the luminosity function
(LF) at rest-frame ultraviolet wavelengths (i.e., the number
density of galaxies as a function of ultraviolet luminosity).
Besides providing information on the number density of galaxies,
the LF can be used to probe the star formation activity in the
universe, because ultraviolet luminosity is sensitive to star
formation. Thus the ultraviolet luminosity density derived
by integrating the LF is related to the star formation rate (SFR)
density in the universe.
By obtaining the LF at various redshifts and examining its evolution,
one can gain insights into the formation history of galaxies and the
star formation history in the universe.

\citet{steidel1999} derived the LF at $z\sim 3$ from their large LBG
sample, finding that their data are well fitted by a Schechter function
with a low-luminosity slope $\alpha$ of $-1.6$ down to $L\simeq 0.1L^\ast$.
They extended their LBG search to $z\sim 4$, to detect no significant
evolution of the LF at bright magnitudes
($M<-21.1$)
from $z\sim 4$ to $z\sim 3$.
More recent observations have also derived the LF at $z\sim 4$ and
$z\sim 5$ \citep{iwata2003, ouchi2004, gabasch2004, sawicki2006}.
The frontier of LBG searches now extends beyond $z\sim 5$.
\citep{yan2003, stanway2003, bunker2004, bouwens2004, bouwens2005,
shimasaku2005}.

However, it is not easy to construct a large sample of LBGs 
beyond $z\sim 4$ covering a wide range of absolute magnitude 
because of their apparent faintness and low surface number 
density.
In the first place, most of the studies to date are limited to bright
LBGs.
Consequently, the shape of the faint end of the LF has not been
well determined.
Thus, the estimation of the cosmic SFR density has a
considerably large uncertainty (about a factor of 3 -- 10) due to
the long extrapolation of the observed LFs to faint magnitudes
\citep[see][]{ouchi2004}.
Exploring LBGs down to faint luminosities is necessary in order to
determine the overall shape of the LF, and to measure the cosmic SFR
density accurately.
Although some surveys are extremely deep, they are restricted to very
small areas (e.g. $\sim 4$ arcmin$^2$ for the Hubble Deep Field, and
$\sim 10$ arcmin$^2$ for the Hubble Ultra Deep Field).
Surveys based on such a small area probably suffer from cosmic
variance, i.e., inhomogeneities in the spatial distribution of LBGs.
Although deep and wide surveys of LBGs at $z\sim 4$ have been 
made by \citet{gabasch2004} and \citet{sawicki2006}
very recently, large samples for $z\gtrsim 4$ are still 
very limited. 
It is therefore crucial to construct a new LBG sample 
from a survey of a similar depth and width and derive the LF 
and the cosmic SFR independently.

In this paper, we present a detailed study of LBGs at $z\sim 4$ -- 5
based on the deep and wide-field images obtained in the
Subaru Deep Field (SDF) Project \citep{kashikawa2004}.
The SDF Project is a program conducted by Subaru Observatory
to carry out a deep galaxy survey over a blank field as large as
$\simeq 0.2$ square degree.
Exploiting these unique data, we make the largest samples of LBGs at
$z\sim 4$ and 5, and derive the LFs down to very faint magnitudes;
$M_\mathrm{UV}=-19.2$ at $z\sim 4$
and $M_\mathrm{UV}=-20.3$ at $z\sim 5$.
This extends the LBG search by \citet{ouchi2004}
based on shallower and smaller area data of the same field
some $\sim 0.5$ mag further
and $1.5$ times wider,
leading to a $\sim 3$ times increase in number of LBGs detected.
These samples enable us to examine the behavior of the LF more
accurately over a wide magnitude range
and obtain more reliable measurements of the cosmic
SFR density in the early universe.
The LFs of LBGs at $z \sim 4$ and 5 we derive are found 
to differ from those at the same redshift ranges 
obtained in some of the previous surveys.

Recently, the LF at ultraviolet wavelengths of present-day
galaxies ($z\sim 0$)
has been derived very accurately from a large galaxy survey
by Galaxy Evolution Explorer (GALEX) \citep{wyder2005}.
\citet{arnouts2005} have derived the LFs at $0.2\le z\le 1.2$ also
based on GALEX observations.
Measurements at $z\sim 6$ are now available
\citep{bouwens2005, shimasaku2005}.
Combining the results at $z\sim 4$ and 5 from this work with those at
lower redshifts and $z\sim 6$ from the literature,
we investigate the evolutionary history of galaxies in terms of the
star formation activity over $0\le z\lesssim 6$.

The outline of this paper is as follows.
Section \ref{sec:data} describes the data of the
SDF Project used in this study.
Section \ref{sec:lbg} describes the selection of LBGs at $z\sim 4$
and 5 from the photometric data.
The contamination by interlopers and the completeness of the samples
are also estimated.
In Section  \ref{sec:lf}, the rest-frame UV luminosity functions
of LBGs at $z\sim 4$ and 5 are derived,
and compared with those by other authors.
In Section \ref{sec:dis} we discuss the evolution of the cosmic
star formation activity over the redshift range $0\le z\le 6$.
The evolution of the luminosity function
and the cosmic star formation rate density is examined.
Section \ref{sec:sum} summarizes this study.

Throughout this paper, the photometric system is based on AB magnitude
\citep{oke1983}.
The cosmology adopted is a flat universe with
$\mathrm{\Omega}_m = 0.3$, $\mathrm{\Omega}_\mathrm{\Lambda} = 0.7$,
and a Hubble constant of $H_0 = 70$ km s$^{-1}$ Mpc$^{-1}$.
These values are consistent with those obtained from the latest CMB
observations \citep{spergel2003}.

\section{DATA} \label{sec:data}

The data used in this study are obtained in the
SDF Project \citep{kashikawa2004}.
The program consists of very deep multi-band optical imaging, NIR
imaging for smaller portions of the field, and follow-up optical
spectroscopy.
This study is based on the optical imaging data together with
additional information from the spectroscopic data.
Full details of the optical imaging observations, data reduction, and
object detection and photometry are described in \citet{kashikawa2004}.

\subsection{Imaging Data} \label{subsec:data-imaging}

The optical imaging observations in the SDF Project
were carried out with the
prime-focus camera \citep[Suprime-Cam:][]{miyazaki2002} mounted on the
8.2 m Subaru Telescope \citep{iye2004} in 2002 and 2003.
The imaging was made for a single field of view of the Suprime-Cam
($34' \times 27'$)
toward the Subaru Deep Field \citep[SDF:][]{maihara2001},
centered on [$13^h24^m38.^s9$, $+27^\circ29'25.''9$ (J2000)].
The scale of the Suprime-Cam is $0.''202$ pixel$^{-1}$.
The images were taken in five standard broad-bands filters: $B$, $V$,
$R$, $i'$, and $z'$,
and two narrow-bands filters: NB816 and NB921.
The central wavelengths of the broad-bands are 4458 \AA, 5478 \AA,
6533 \AA, 7684 \AA, and 9037 \AA\ for $B$, $V$, $R$, $i'$, and $z'$,
respectively.
The SDF was previously imaged during the commissioning runs
of the Suprime-Cam in 2001 in five bandpasses,
$B$, $V$, $R$, $i'$, and $z'$, with 1 -- 3 hours exposure times.
The work by \citet{ouchi2004} is based on these data.
They were co-added in constructing the final images.

All individual CCD data with good quality were reduced
and combined to make a single
image for each band using a pipeline software package
\citep{ouchi2003} whose core programs were taken from IRAF and
mosaic-CCD data reduction software \citep{yagi2002}.
The combined images for all seven bands were aligned
and smoothed with Gaussian kernels
so that all have the same seeing size,
a PSF FWHM of $0.''98$.
The efficiency and reliability of object detection and photometry
in regions near the edges of the images are significantly lower
on account of low $S/N$ ratios caused by dithering.
Regions around very bright stars are also degraded due to bright haloes
and saturation trails.
We carefully defined these low-quality regions, and did not use objects
located in these regions.
The effective area of the final images
after removal of the low-quality regions
is 875 arcmin$^2$.
Photometric calibration for the images was made using
standard stars taken during the observations.
The total exposure time reaches $\sim 10$ hours for each band, and
the limiting magnitude ($3\sigma$ within a $2''$ diameter aperture)
is 28.45 ($B$), 27.74 ($V$), 27.80 ($R$), 27.43 ($i'$), 26.62 ($z'$),
26.63 (NB816), and 26.54 (NB921).

Object detection and photometry were performed using SExtractor version
2.1.6 \citep{bertin1996}.
A collection of at least 5 contiguous pixels above the $2\sigma$ sky
noise was identified as an object.
The object detection was made for all seven images independently.
For each of the objects detected in a given band image, photometry was
made in all the images at exactly the same position as
in the detection-band image.
Thus, seven catalogs were constructed for respective detection bands.
For the present study, we limit the catalogs to objects whose
detection-band magnitudes are brighter than the $5\sigma$ limiting
magnitude ($5\sigma$ sky noise within a $2''$ diameter aperture) of
that band, in order to provide a reasonable level of photometric
completeness.
We use magnitudes within a $2''$ diameter aperture to derive the
colors of objects,
and adopt MAG\_AUTO in SExtractor for total magnitudes.
The magnitudes of objects were corrected for a small amount of
foreground Galactic extinction toward the SDF using the dust map of
\citet{schlegel1998}.
Since the variation of $E(B-V)$ over the SDF of 875 arcmin$^2$ 
is at most 0.007, we adopt a single value, $E(B-V)=0.017$, 
for correction, which is the value at the center of the SDF. 
The value of $E(B-V)=0.017$ corresponds to an extinction of 
$A_B = 0.07$, $A_V = 0.05$, $A_R = 0.04$, $A_{i'} = 0.03$,
$A_{z'} = 0.02$, $A_\mathrm{NB816} = 0.03$, and
$A_\mathrm{NB921} = 0.02$.

\subsection{Spectroscopic Data} \label{subsec:data-spectroscopy}

We carried out spectroscopic follow-up observations
for some objects in our catalogs
with the Subaru Faint Object Camera and Spectrograph
\citep[FOCAS:][]{kashikawa2002} on the Subaru Telescope
during 2002 -- 2004
and DEIMOS \citep{faber2003} on the Keck II Telescope in 2004.
The spectroscopic observations were made in
the multi-object slit mode.
For the FOCAS observations,
we used a 300 line mm$^{-1}$ grating with an O58 order-cut filter
and an Echelle with a $z'$ filter.
Slit widths on the masks were either $0.''8$ or $0.''6$
for the 300 line configuration,
resulting in a spectral resolution
of 9.5 \AA\ and 7.1 \AA\ at 8150 \AA, respectively,
and $0.''8$ for the Echelle.
The integration times of respective masks were 7000 -- 9000 seconds.
Flux calibration was made with spectra of the spectroscopic standard
stars Hz 44 and Feige 34.
The data were reduced in a standard manner.
For the observations with DEIMOS,
we used a 830 line mm$^{-1}$ grating with a GG495 order-cut filter.
Slit widths were $1.''0$,
which gave a spectral resolution
of 3.97 \AA.
The integration times were 6300 -- 19800 seconds.
Spectra of the standard stars BD+28d4211 and Feige 110 were taken
for flux calibration.
The data were reduced with the spec2d pipeline
\footnote{The spec2d pipeline was developed at UC Berkeley
with support from NSF grant AST-0071048.}
for the reduction of DEEP2 DEIMOS data.
In addition, we also use spectra of objects in the SDF
which were taken during the guaranteed time observations
of the FOCAS in 2001 \citep{kashikawa2003,ouchi2004}.

\begin{figure}[t]
  \begin{center}
    \includegraphics[scale=0.4]{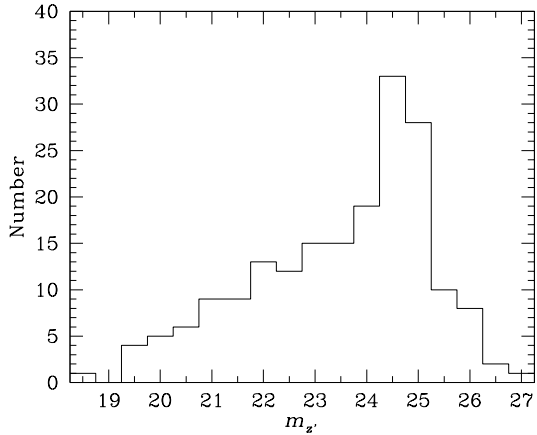}
  \end{center}
  \caption{%
The magnitude distribution of the spectroscopic objects
with $z_\mathrm{spec}<5.5$ in our catalogs.
\label{fig:nm}}
\end{figure}

\begin{figure}[t]
  \begin{center}
    \includegraphics[scale=0.4]{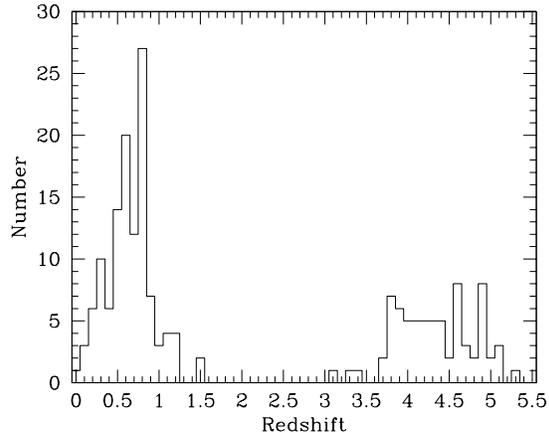}
  \end{center}
  \caption{%
The redshift distribution of the spectroscopic objects
with $z_\mathrm{spec}<5.5$ in our catalogs.
\label{fig:nz}}
\end{figure}

In Figures \ref{fig:nm} and \ref{fig:nz}, we present
the distribution of the $z'$ magnitude of the spectroscopic objects
and their redshift distribution, respectively.
Only objects with $z_\mathrm{spec}<5.5$ are included here.

\section{LYMAN-BREAK GALAXY SAMPLES AT $z\sim 4$ AND $5$}
\label{sec:lbg}

\subsection{Selection of Lyman-break Galaxies}
\label{subsec:lbg-select}

\begin{figure}[t]
  \begin{center}
    \includegraphics[scale=0.4]{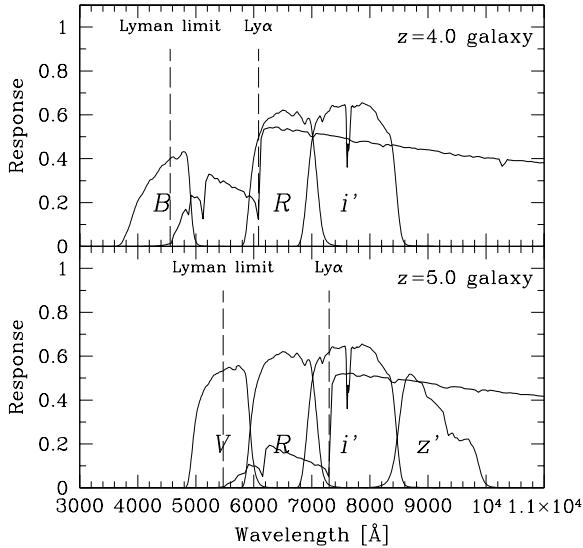}
  \end{center}
  \caption{%
\textit{Upper panel}: The $B$, $R$, and $i'$ bandpasses
overplotted on the spectrum of a generic $z=4$ galaxy
(thick line), illustrating the utility of color
selection technique using these three bandpasses for
locating $z\sim 4$ galaxies.
\textit{Lower panel}: The $V$, $R$, $i'$, and $z'$ bandpasses
overplotted on the spectrum of a generic $z=5$ galaxy
(thick line).
The band sets of $Vi'z'$ and $Ri'z'$ work well to
isolate $z\sim 5$ galaxies.
\label{fig:sed}}
\end{figure}

We demonstrate the principle of the Lyman-break color selection
technique to locate galaxies at $z\sim 4$ and $z\sim 5$
in \reffig{sed}.
For galaxies at $z\sim 4$,
the break in the spectrum at rest-frame ultraviolet wavelength
enters the $B$ band
and the flat continuum longward of Lyman-$\alpha$ shifts to
wavelengths longer than the $R$ band, and these galaxies are
identified by their red \br\ color and blue \ri\ color.
We select LBGs at $z\sim 4$ by these two colors (hereafter called
$BRi'$-LBGs).
For galaxies at $z\sim 5$, the spectral break enters into the $V$ band
and the flat continuum is sampled at wavelengths longer than the $i'$
band.
At $z\sim 5$, the attenuation due to the Lyman-$\alpha$ forest is
so strong that the flux in the $R$ band is severely suppressed, as
well as the flux in the $V$ band.
Thus, galaxies at $z\sim 5$ are identified by their red \vi\ or
\ri\ color and blue \iz\ color.
We make two samples for LBGs at $z\sim 5$: one selected by \vi\ and
\iz\ colors, and the other selected by \ri\ and \iz\ colors
(hereafter called $Vi'z'$-LBGs and $Ri'z'$-LBGs, respectively).

\begin{figure}[h]
  \begin{center}
    \includegraphics[scale=0.5]{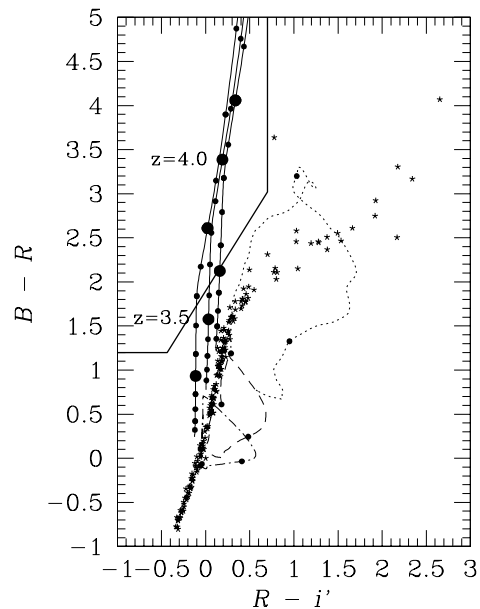}
  \end{center}
  \caption{%
\br\ vs \ri\ diagram where the predicted colors of model
galaxies and stars are illustrated.
The three solid lines indicate the tracks for model spectra of
young star-forming galaxies with reddening of $E(B-V)=0$, $0.16$,
and $0.3$ (from left to right).}
The redshift range is from $z=3$ to higher redshifts,
and the circles on the track mark the redshift interval of 0.1.
The dotted, dashed, and dot-dashed lines delineate the tracks for
model spectra of local elliptical, spiral, and irregular galaxies,
respectively, redshifted from $z=0$ to $3$ without evolution.
The circles on each track mark $z=0$, $1$, and $2$.
The asterisks represent the colors of 175 Galactic stars
given by \citet{gunn1983}.
The thick line indicates the boundary which we adopt for
the selection of $BRi'$-LBGs.
\label{fig:model-bri}
\end{figure}

\begin{figure}[t]
  \begin{center}
    \includegraphics[scale=0.5]{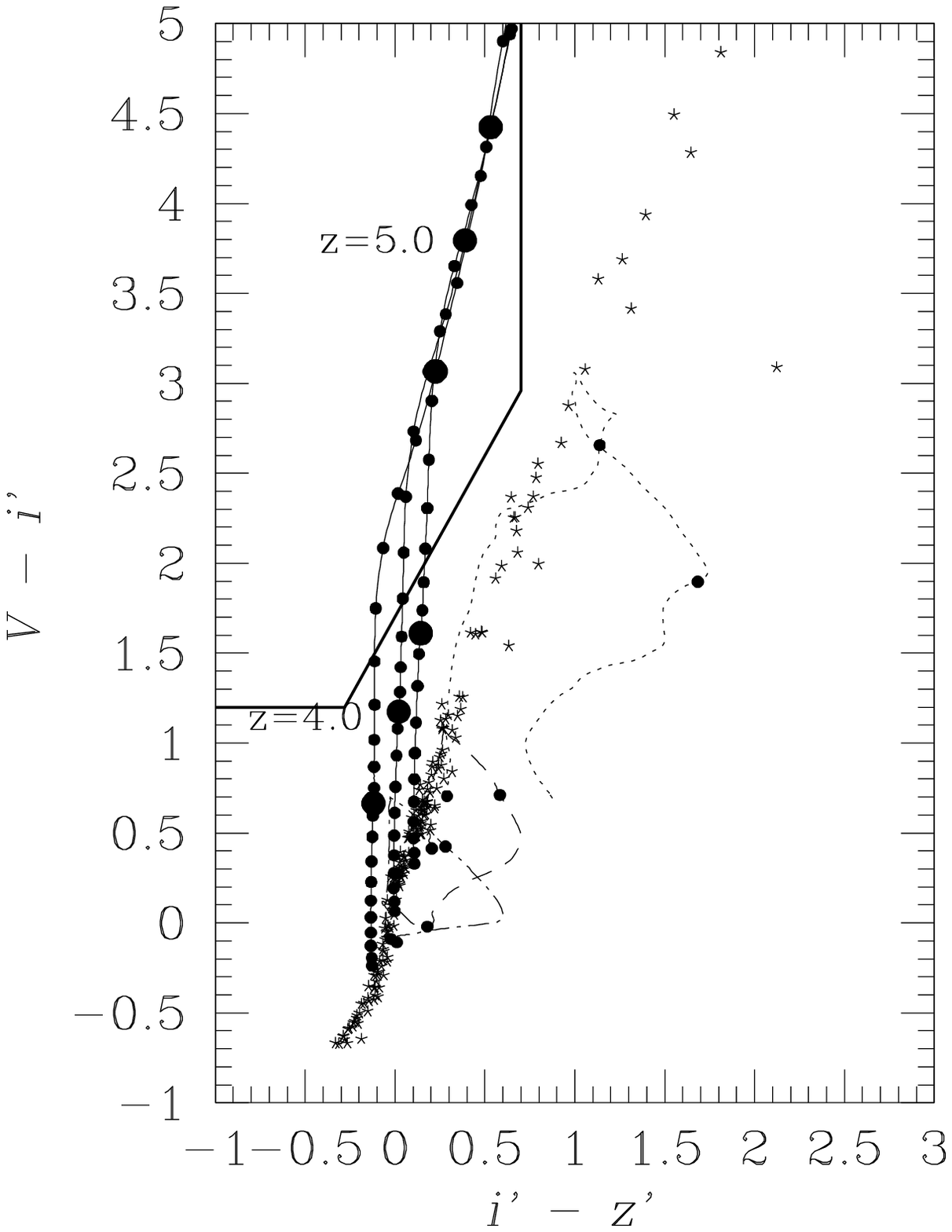}
  \end{center}
  \caption{%
\vi\ vs \iz\ diagram where the predicted colors of model galaxies
and stars are illustrated.
Lines and symbols are the same as in \reffig{model-bri}.
The thick line indicates the boundary which we adopt for
the selection of $Vi'z'$-LBGs.
\label{fig:model-viz}}
\end{figure}

\begin{figure}[t]
  \begin{center}
    \includegraphics[scale=0.5]{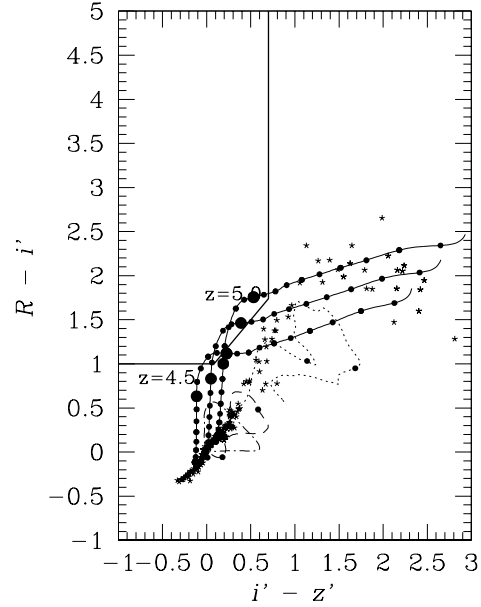}
  \end{center}
  \caption{%
\ri\ vs \iz\ diagram where the predicted colors of model galaxies
and stars are illustrated.
Lines and symbols are the same as in \reffig{model-bri}.
The thick line indicates the boundary which we adopt for
the selection of $Ri'z'$-LBGs.
\label{fig:model-riz}}
\end{figure}

Figures \ref{fig:model-bri} -- \ref{fig:model-riz} illustrate the
predicted positions of high-redshift galaxies and foreground
objects (lower-redshift galaxies and Galactic stars) in three
two-color diagrams: \br\ vs \ri, \vi\ vs \iz, and
\ri\ vs \iz.
The solid lines in these figures indicate the tracks for model
spectra of young star-forming galaxies for the redshift range $z>3$.
The model spectrum was constructed using
the stellar population synthesis
code developed by \citet{kodama1997}.
For model parameters, an age of 0.1 Gyr, a Salpeter initial mass
function, and a star-formation timescale of 5 Gyr
were adopted, and reddening of $E(B-V) = 0.16$ was applied using the
dust extinction formula for starburst galaxies by
\citet{calzetti2000}.
These values reproduce the average rest-frame ultraviolet-optical
spectral energy distribution
of LBGs observed at $z\sim 3$ \citep{papovich2001}.
We also show the tracks for model spectra with
reddening of $E(B-V) = 0$ and $0.3$ for reference.
The absorption due to the intergalactic medium was applied following the
prescription by \citet{madau1995}.
The dotted, dashed, and dot-dashed lines delineate the tracks for model
spectra of local elliptical, spiral, and irregular galaxies,
respectively, redshifted from $z=0$ to $z=3$.
These spectra were also constructed using the stellar population
synthesis code by \citet{kodama1997}, and redshifted without
evolution.
The asterisks denote the 175 Galactic stars of
various types whose spectra are given by \citet{gunn1983}.
The colors are calculated by convolving
constructed model spectra of galaxies
or given spectra of stars with the response functions of the
Suprime-Cam filters.
It can be seen that the model young star-forming galaxy moves
nearly vertically from the origin in these two-color diagrams at $z>3$,
namely it becomes redder in
\br , \vi , and \ri\ colors, while keeping \ri , \iz , and \iz colors
blue, as redshift increases, and that it shifts into a portion
of the two-color space which is unpopulated by foreground objects.
At $z>5$, the \ri\ color does not become redder very much while \iz\ 
color becomes redder in the \ri\ vs \iz\ diagram.
These figures imply that LBGs at $z\sim 4$ and those at
$z\sim 5$
can be well isolated from foreground objects in the \br\ vs \ri\ 
diagram, and in the \vi\ vs \iz\ and \ri\ vs \iz\ diagrams,
respectively.
The most critical contamination is caused by elliptical galaxies at
$z\sim 0.5$, which come close to high-redshift galaxies in these
diagrams due to the 4000 \AA\ break on their spectra.

\begin{figure}[t]
  \begin{center}
    \includegraphics[scale=0.35]{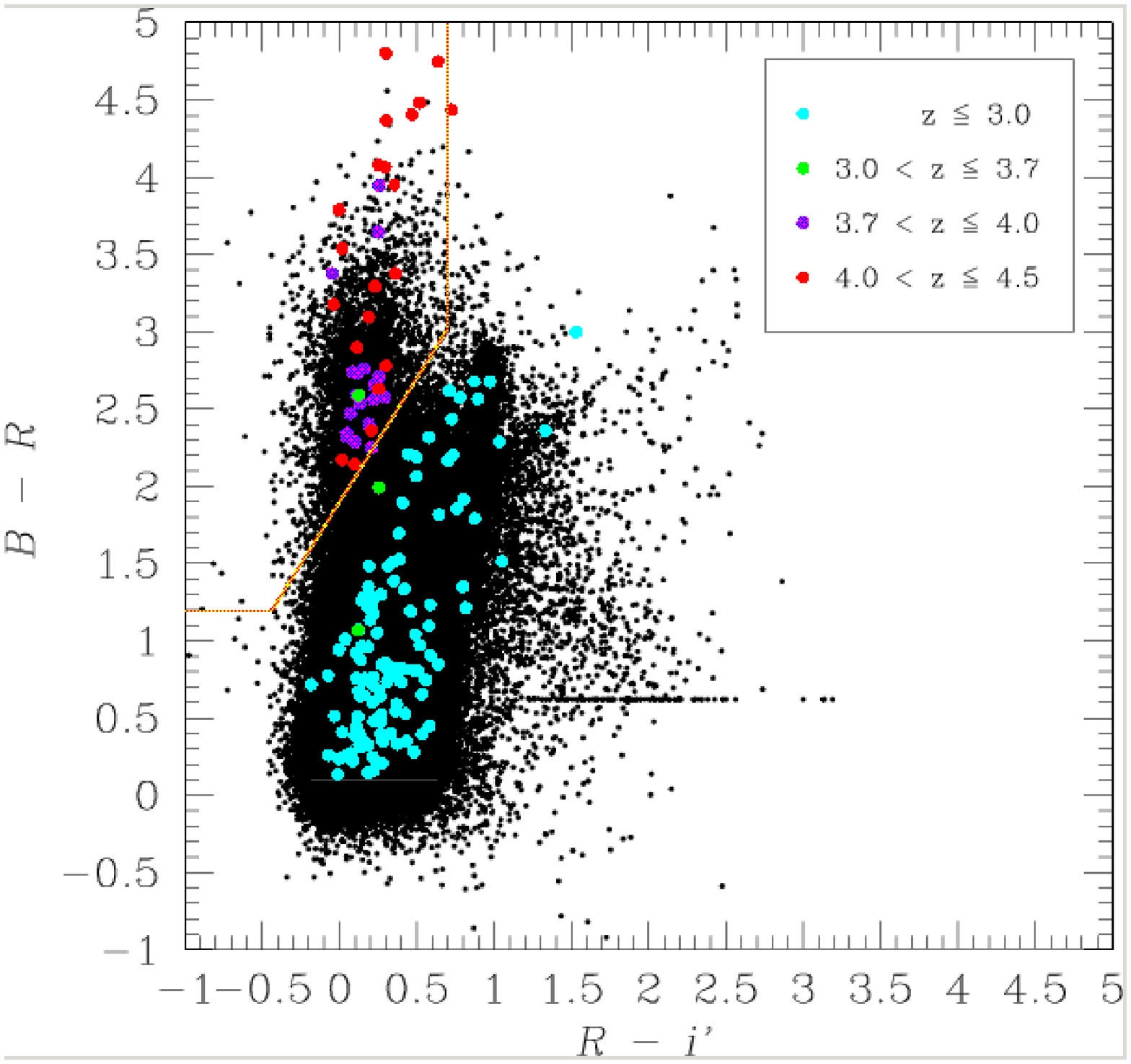}
  \end{center}
  \caption{%
\br\ vs \ri\ diagram for all the objects contained in the
$i'$-detection catalogs (106,025 objects down to $i'\le 26.85$).
When the $B$ and/or $R$ magnitude of an object is fainter
than the $1\sigma$ magnitude of the band, the $1\sigma$
magnitude is assigned to the object.
The horizontal sequence along $\br = 0.62$ shows objects
which are fainter than the $1\sigma$ magnitude in both $B$ and $R$.}
The colored symbols show objects with spectroscopic redshifts,
where cyan, green, violet, and red represent objects
in the range
$z<3.0$, $3.0\le z<3.7$, $3.7\le z<4.0$, and $4.0\le z<4.5$,
respectively.
The thick orange line indicates the boundary which we adopt for
the selection of $BRi'$-LBGs.
\label{fig:detect-bri}
\end{figure}

\begin{figure}[t]
  \begin{center}
    \includegraphics[scale=0.35]{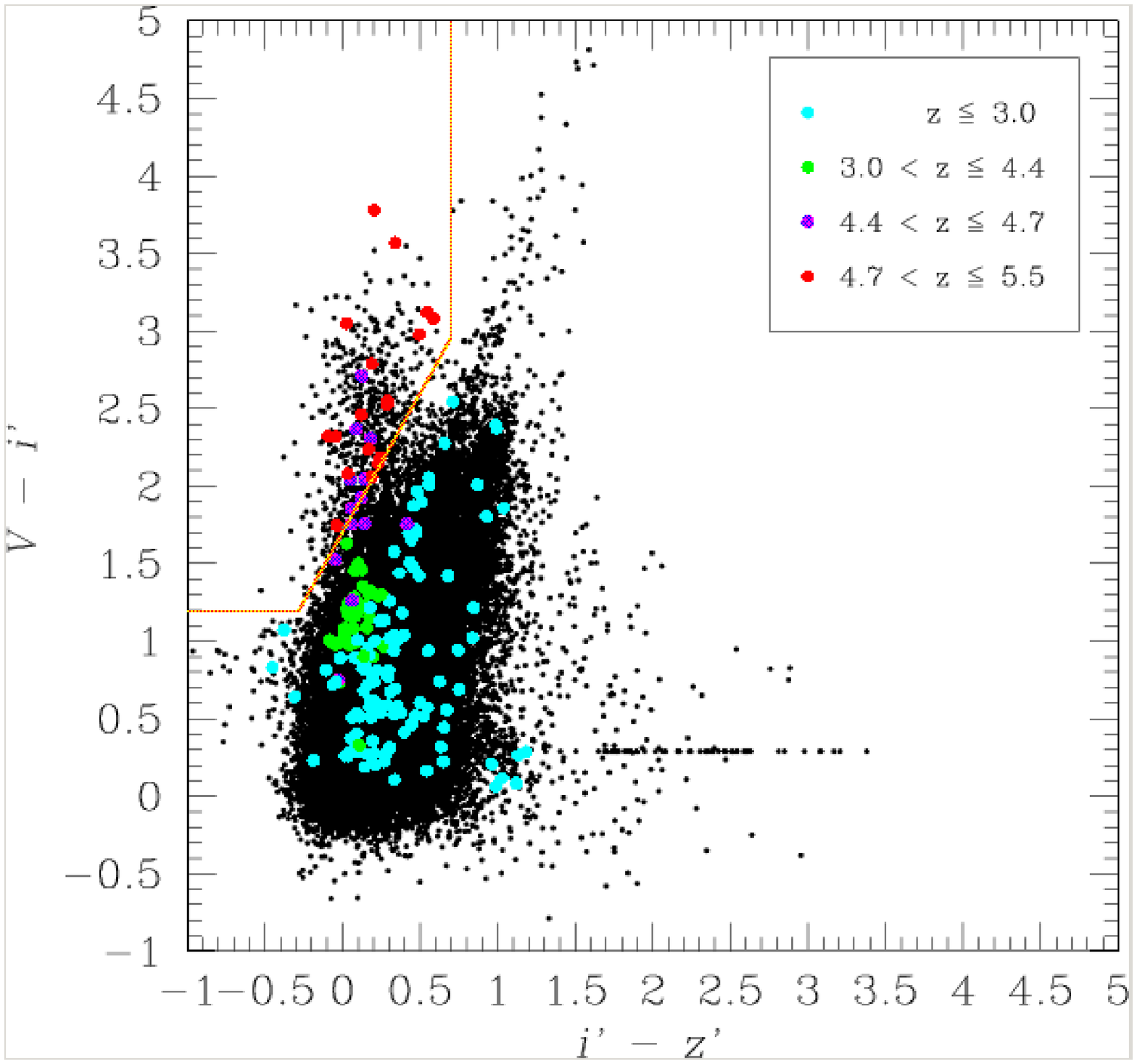}
  \end{center}
  \caption{%
\vi\ vs \iz\ diagram for all the objects contained in the
$z'$-detection catalogs (76,351 objects down to $z'\le 26.05$).
When the $V$ and/or $i'$ magnitude of an object is fainter
than the $1\sigma$ magnitude of the band, the $1\sigma$
magnitude is assigned to the object.
The horizontal sequence along $\vi = 0.29$ shows objects
which are fainter than the $1\sigma$ magnitude in both $V$ and $i'$.}
The colored symbols show objects with spectroscopic redshifts,
where cyan, green, violet, and red represent objects
in the range
$z<3.0$, $3.0\le z<4.4$, $4.4\le z<4.7$, and $4.7\le z<5.5$,
respectively.
The thick orange line indicates the boundary which we adopt for
the selection of $Vi'z'$-LBGs.
\label{fig:detect-viz}
\end{figure}

\begin{figure}[t]
  \begin{center}
    \includegraphics[scale=0.35]{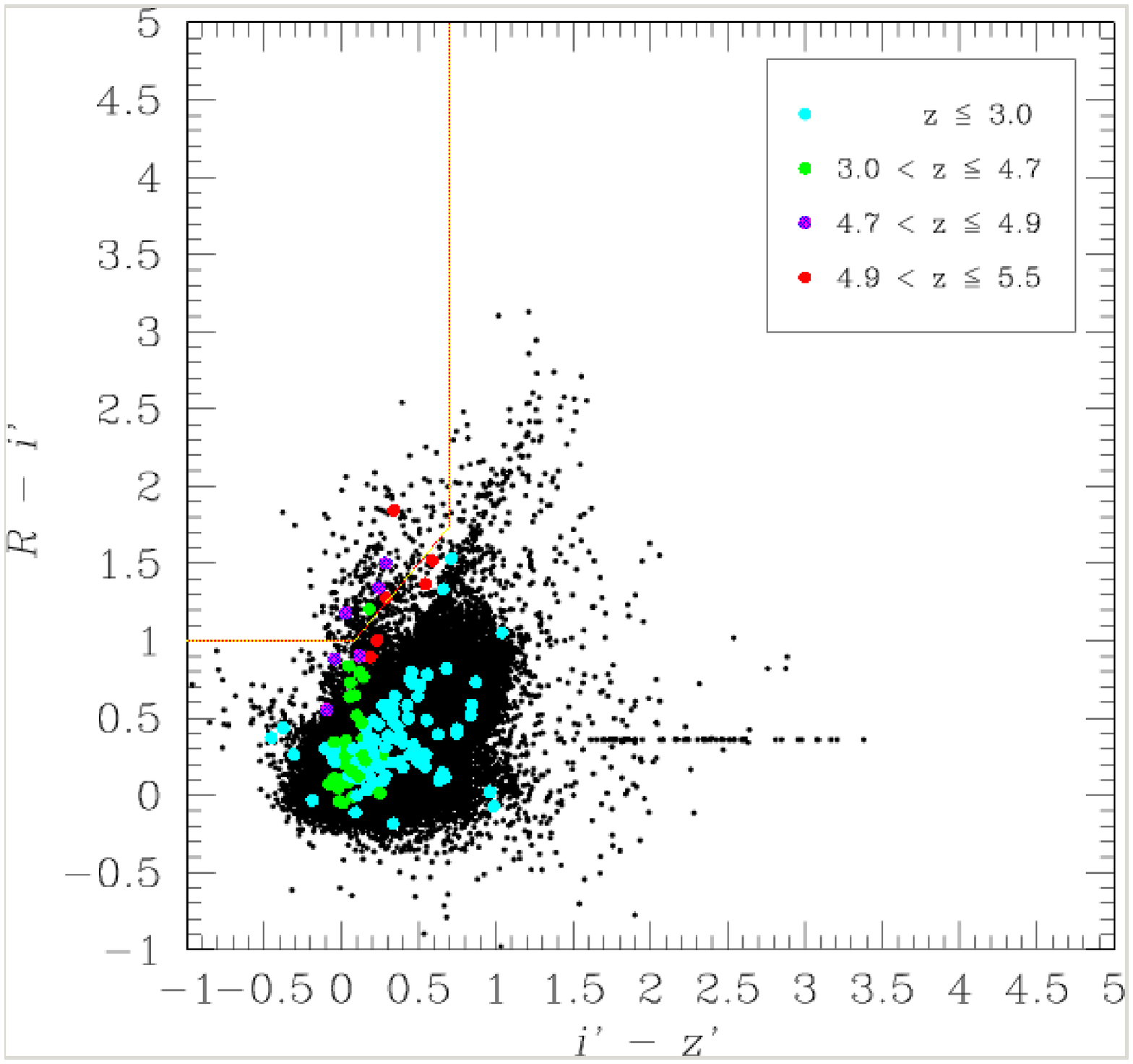}
  \end{center}
  \caption{%
\ri\ vs \iz\ diagram for all the objects contained in the
$z'$-detection catalogs (76,351 objects down to $z'\le 26.05$).
When the $R$ and/or $i'$ magnitude of an object is fainter
than the $1\sigma$ magnitude of the band, the $1\sigma$
magnitude is assigned to the object.
The horizontal sequence along $\ri = 0.35$ shows objects
which are fainter than the $1\sigma$ magnitude in both $R$ and $i'$.}
The colored symbols show objects with spectroscopic redshifts,
where cyan, green, violet, and red represent objects
in the range
$z<3.0$, $3.0\le z<4.7$, $4.7\le z<4.9$, and $4.9\le z<5.5$,
respectively.
The thick orange line indicates the boundary which we adopt for
the selection of $Ri'z'$-LBGs.
\label{fig:detect-riz}
\end{figure}

In Figures \ref{fig:detect-bri} -- \ref{fig:detect-riz}, we show the
distribution of all the objects contained in the catalogs in the
two-color diagrams.
When the magnitude of an object in a non-detection band is fainter
than the $1\sigma$ magnitude of the band, the $1\sigma$ magnitude is
assigned to the object.
Objects with spectroscopic redshifts are shown with colored
symbols in these figures, where different colors mean different
redshift bins.
It is found that the colors of spectroscopically identified objects
match fairly well with those of model galaxies with corresponding
redshifts in Figures \ref{fig:model-bri} -- \ref{fig:model-riz}.

We adopt the same color criteria for $BRi'$-LBGs and $Vi'z'$-LBGs as
those of \citet{ouchi2004}.
For $Ri'z'$-LBGs, we visually fine-tune their color criteria based on
the increased redshift information.
Specifically, we set the selection criteria for $BRi'$-LBGs as:
\begin{eqnarray}
  && B-R > 1.2, \nonumber\\
  && R-i' < 0.7, \\
  && B-R > 1.6(R-i') + 1.9, \nonumber
\end{eqnarray}
for $Vi'z'$-LBGs as:
\begin{mathletters}
  \begin{eqnarray}
    && V-i' > 1.2, \nonumber\\
    && i'-z' < 0.7, \\
    && V-i' > 1.8(i'-z') + 1.7, \nonumber\\
    && \nonumber\\
    && B > 3\sigma\ \mathrm{mag},
  \end{eqnarray}
\end{mathletters}
and for $Ri'z'$-LBGs as:
\begin{mathletters}
  \begin{eqnarray}
    && R-i' > 1.0, \nonumber\\
    && i'-z' < 0.7, \\
    && R-i' > 1.2(i'-z') + 0.9, \nonumber\\
    && \nonumber\\
    && B,\ V > 3\sigma\ \mathrm{mag}.
  \end{eqnarray}
\end{mathletters}
The boundaries on the two-color diagrams defined by these color criteria
are outlined with the thick orange lines in Figures
\ref{fig:model-bri} -- \ref{fig:model-riz} and
\ref{fig:detect-bri} -- \ref{fig:detect-riz}.
For the bands which are placed entirely shortward of the Lyman break,
we additionally require that LBGs should be non-detected ((2b) and
(3b)), which is expected to work effectively to reduce contamination
from foreground galaxies.

We apply these selection criteria to the catalogs.
We use the $i'$-detection catalogs for $BRi'$-LBGs, which contain
$106,025$ objects to a limit of $i'=26.85$ (total magnitude corrected
for the Galactic extinction), and the $z'$-detection catalogs for
$Vi'z'$-LBGs and $Ri'z'$-LBGs, which contain $76,351$ objects
to $z'=26.05$
(total magnitude corrected for the Galactic extinction).
The number of objects which satisfy the selection criteria, i.e., the
number of LBG candidates, is $3,808$ for $BRi'$-LBGs, 539 for
$Vi'z'$-LBGs, and 240 for $Ri'z'$-LBGs
\footnote{For this study, we limit the catalogs to the 5$\sigma$
limiting magnitude.
\citet{kashikawa2006} also use our LBG samples for their study
on clustering properties of LBGs but down to the 3$\sigma$
limiting magnitude.
The number of LBG candidates with magnitudes brighter than
the 3$\sigma$ limiting magnitude is $4,543$ for $BRi'$-LBGs,
831 for $Vi'z'$-LBGs, and 386 for $Ri'z'$-LBGs.
This difference causes no significant effect on luminosity functions
except for the faintest bins.}.
\reftbl{sample} summarizes the samples of the LBG candidates.

\begin{figure}[h]
  \begin{center}
    \includegraphics[scale=0.5]{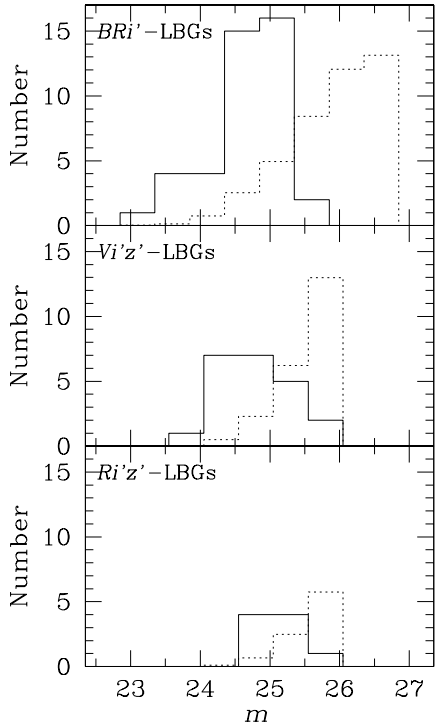}
  \end{center}
  \caption{%
The magnitude distribution of the spectroscopically identified
objects in each LBG sample.
Dotted histograms show the magnitude distribution of all the
objects in each LBG sample normalized to the number of the
spectroscopically identified objects.
\label{fig:spec-nm}}
\end{figure}

\begin{figure}[h]
  \begin{center}
    \includegraphics[scale=0.5]{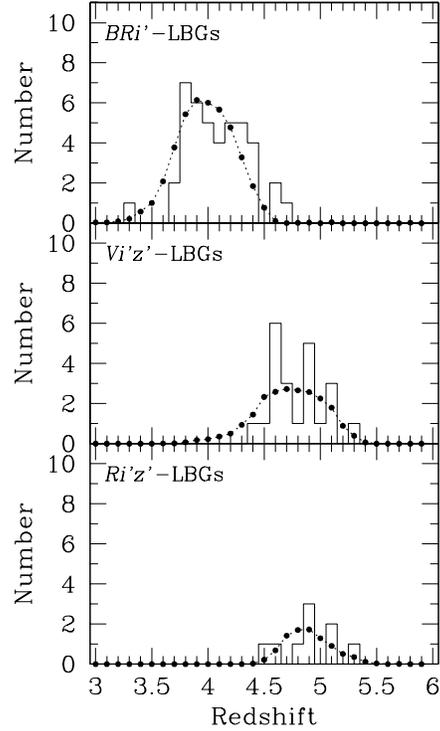}
  \end{center}
  \caption{%
The redshift distribution of the spectroscopically identified
objects in each LBG sample.
Dotted lines show the redshift distribution functions from the
Monte Carlo simulation (see \S \ref{subsec:lbg-comp})
normalized to the number of the spectroscopically
identified objects.
\label{fig:spec-nz}}
\end{figure}

\begin{figure}[t]
  \begin{center}
    \includegraphics[scale=0.4]{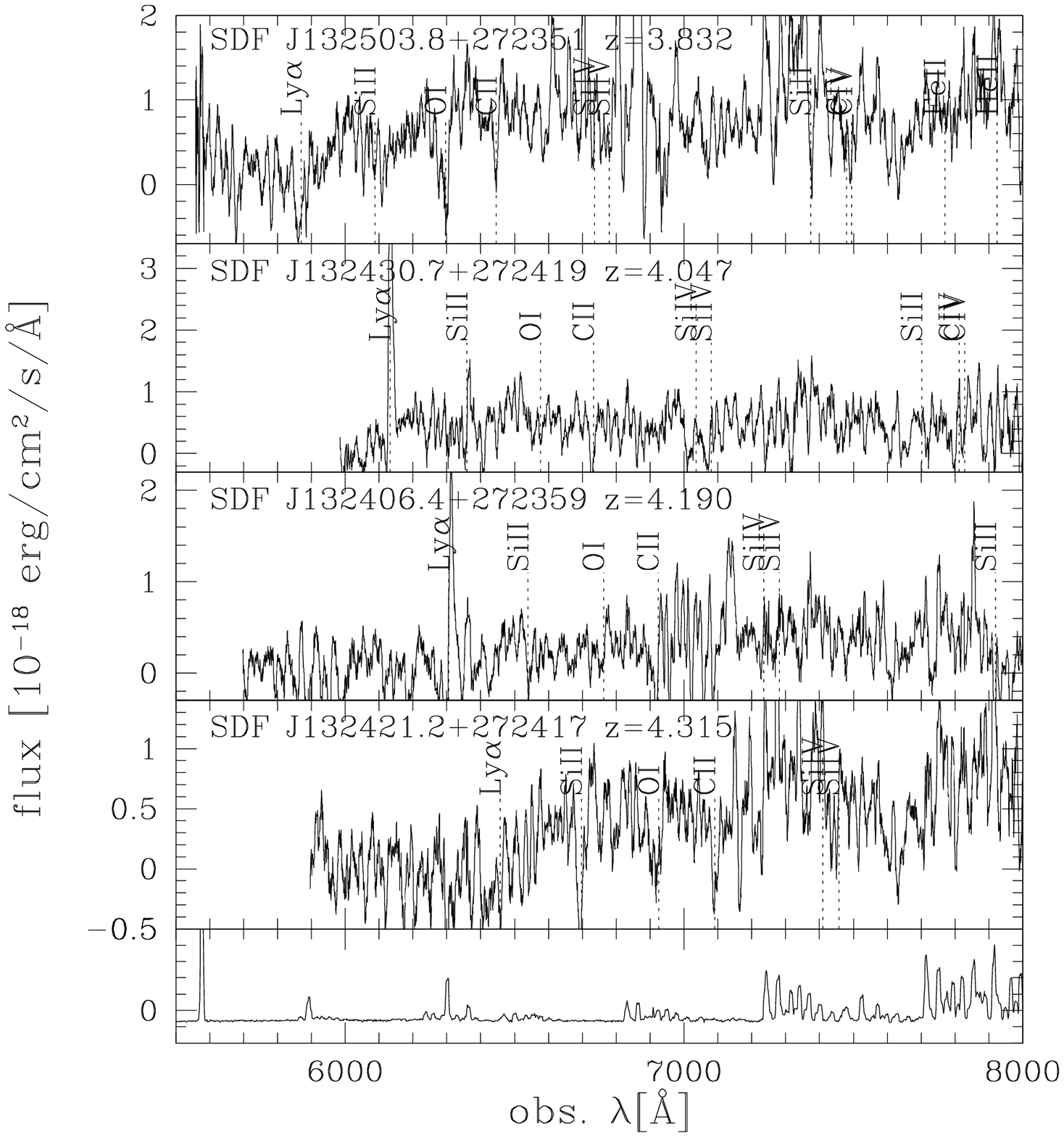}
  \end{center}
  \caption{%
Examples of spectra of Lyman-break galaxies at $z\sim 4$
with good quality that have prominent spectral features
such as a strong Ly$\alpha$ emission line and interstellar lines.}
The lowest panel shows a relative night-sky spectrum.
\label{fig:spec-z4}
\end{figure}

\begin{figure}[t]
  \begin{center}
    \includegraphics[scale=0.4]{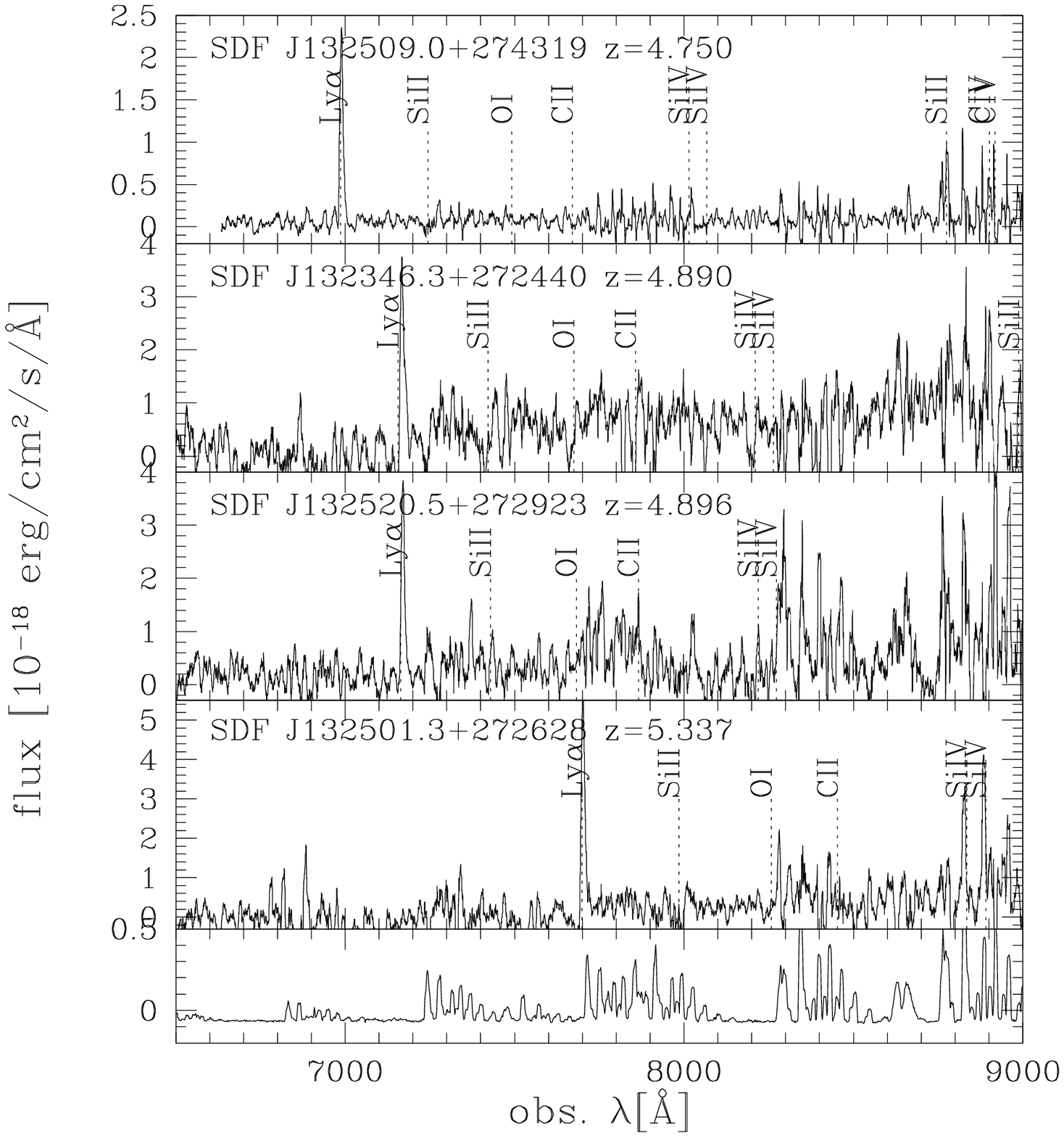}
  \end{center}
  \caption{%
Same as \reffig{spec-z4}, but for $z\sim 5$.
\label{fig:spec-z5}}
\end{figure}

The selection criteria are shown by our spectroscopy to be
fairly reliable.
We spectroscopically observed 105 objects in the $BRi'$-LBG sample,
and identified 42 objects.
Among the 42, only one is an interloper at $z=3.29$,
with the remaining 41 being at $z\ge 3.7$.
Similarly, 32 objects from the $Vi'z'$-LBG sample
were spectroscopically observed, and 23 were identified.
All but one of the 23 objects are at $z\ge 4.4$;
the remaining one appears to be a Galactic star.
In the $Ri'z'$-LBG sample, 12 objects were spectroscopically observed,
and 9 were identified.
All are found to be at $z\ge 4.5$.
Possible reasons for the low detection rate for $BRi'$-LBGs are
that the Ly$\alpha$ emission of $z\sim 4$ LBGs is on average weaker
than that of $z\sim 5$ LBGs and that the Ly$\alpha$ line at $z\sim 4$
falls in the wavelength range where the sensitivity of
our spectroscopy is relatively low.
The failed sample may also include interlopers between 
$z \sim 2$ and $3$ whose strong line features do not enter 
the wavelength coverage of our spectrograph.
From the following simulation, however, we infer that 
the fraction of such interlopers is not so high in our LBG sample.
First, we add noise to the colors of all objects outside 
of the $BRi'$ selection criteria in our photometric catalog 
according to their measured magnitudes.
We then count the objects which now meet the selection 
criteria due to the modified colors, to find that they are 
less than $10 \%$ of the original LBG candidates 
even for the faintest magnitude bin where photometric errors 
are the largest.
Note that this simulation gives an upper limit to the fraction 
of interlopers, since the measured colors already include 
photometric errors.
In \reffig{spec-nm}, we present the apparent magnitude distribution
of the spectroscopically identified objects in each LBG sample.
Although the targets of spectroscopy is biased toward
bright objects,
we emphasize here that the spectroscopic samples also include
faint objects and there are no interlopers even among them.
We cannot, however, rule out the possibility that there may
be interlopers for which no redshift have been secured.
\reffig{spec-nz} shows the redshift distribution of the
spectroscopic samples.
The average redshift
is $<z>_\mathrm{spec} = 4.1$ for $BRi'$-LBGs,
$<z>_\mathrm{spec} = 4.8$ for $Vi'z'$-LBGs,
and $<z>_\mathrm{spec} = 4.9$ for $Ri'z'$-LBGs.
Examples of spectra of $z\sim 4$ LBGs and $z\sim 5$ LBGs are shown
in Figures \ref{fig:spec-z4} and \ref{fig:spec-z5}, respectively.

\begin{deluxetable}{cccc} 
  \tablecolumns{4} 
  \tablewidth{0pt} 
  \tablecaption{Samples of LBG candidates \label{tbl:sample}} 
  \tablehead{
    \colhead{\hspace{0.1cm} Sample Name\hspace{0.1cm} }
    & \colhead{\hspace{0.1cm} Number of Candidates\hspace{0.1cm} }
    & \colhead{\hspace{0.1cm} Detection Band\hspace{0.1cm} }
    & \colhead{\hspace{0.1cm} Magnitude Range\tablenotemark{\ast}\hspace{0.1cm} }
  }
  \startdata 
$BRi'$-LBG  & 3,808   & $i'$ & $i'=22.85 - 26.85$ \\
$Vi'z'$-LBG & \phn\phd 539 & $z'$ & $z'=23.55 - 26.05$ \\
$Ri'z'$-LBG & \phn\phd 240 & $z'$ & $z'=24.05 - 26.05$ \\
  \enddata 
  \tablenotetext{\ast\ }{Total magnitudes corrected for the Galactic extinction.}
\end{deluxetable} 

\clearpage

\subsection{Number Counts} \label{subsec:lbg-num}

\begin{figure}[h]
  \begin{center}
    \includegraphics[scale=0.4]{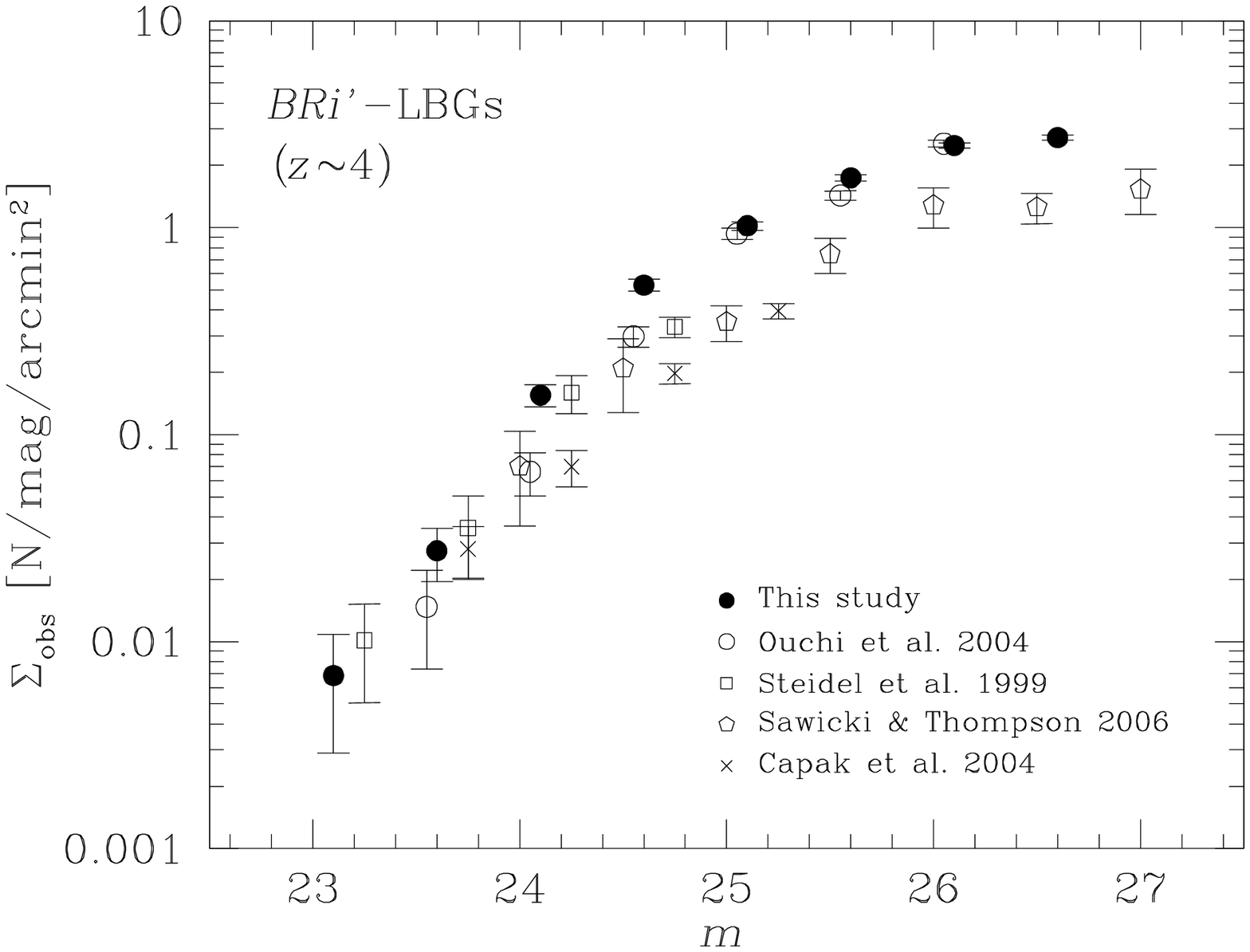}
  \end{center}
  \caption{%
Observed number counts of LBGs at $z\sim 4$ uncorrected for
incompleteness or contamination.
The filled circles represent the $BRi'$-LBGs in this study.
Comparison is made with the previous measurements obtained by
\citet{ouchi2004} (open circles),
\citet{steidel1999} (open squares),
\citet{sawicki2006} (open pentagons),
and \citet{capak2004} (crosses).
The error bars for this study, \citet{ouchi2004},
and \citet{capak2004} reflect Poisson errors.
The error bars for \citet{steidel1999}
and \citet{sawicki2006}
include 
an estimate of cosmic variance from their multiple fields
as well as Poisson errors.
\label{fig:num-bri}}
\end{figure}

\begin{figure}[h]
  \begin{center}
    \includegraphics[scale=0.4]{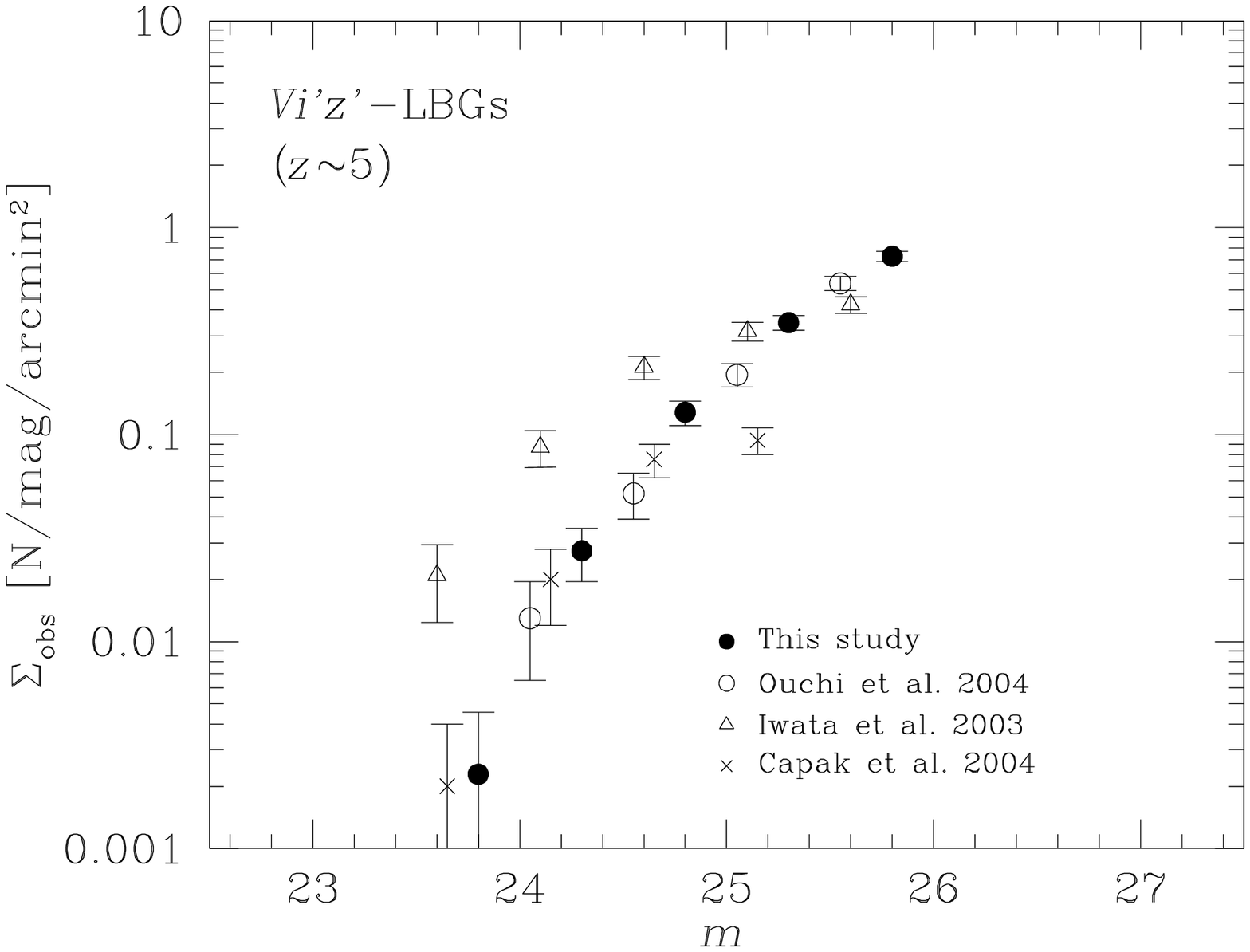}
  \end{center}
  \caption{%
Observed number counts of $V$-dropout LBGs at $z\sim 5$
uncorrected for incompleteness or contamination.
The filled circles represent the $Vi'z'$-LBGs in this study.
Comparison is made with the previous measurements obtained by
\citet{ouchi2004} (open circles),
\citet{iwata2003} (open triangles),
and \citet{capak2004} (crosses).
The error bars reflect Poisson errors.
\label{fig:num-viz}}
\end{figure}

\begin{figure}[h]
  \begin{center}
    \includegraphics[scale=0.4]{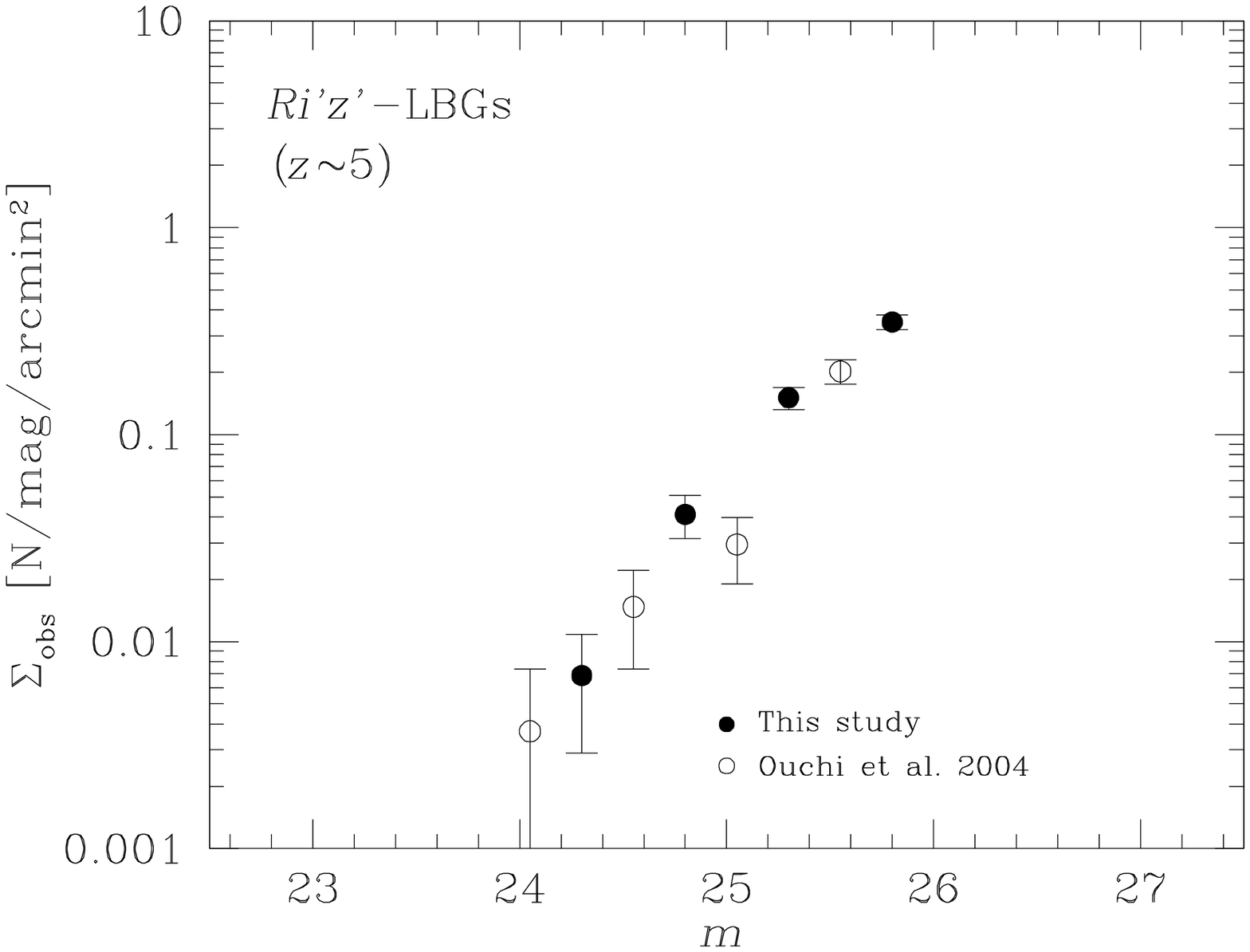}
  \end{center}
  \caption{%
Observed number counts of $R$-dropout LBGs at $z\sim 5$
uncorrected for incompleteness or contamination.
The filled circles represent the $Ri'z'$-LBGs in this study.
Comparison is made with the previous measurements obtained by
\citet{ouchi2004} (open circles).
The error bars reflect Poisson errors.
\label{fig:num-riz}}
\end{figure}

The observed number counts of the LBG candidates as a function of
apparent magnitude are shown in Figures \ref{fig:num-bri} --
\ref{fig:num-riz}, along with those measured by other authors who
selected similar LBGs.
None of the data is corrected for contamination or incompleteness.
Previous LBG surveys included in the figures are summarized in
\reftbl{ref}.
\citet{ouchi2004} used the same photometric systems to make samples of
LBGs at $z\sim 4$ over a 543 arcmin$^2$ area and LBGs at $z\sim 5$
over a 616 arcmin$^2$ area of the SDF.
\citet{steidel1999} carried out a search for LBGs at $z\sim 4$ using
\gr, \rrii\ colors in a total of 828 arcmin$^2$ area of 10 separate
fields.
\citet{sawicki2006} undertake a survey of LBGs at $z\sim 4$
with the same photometric systems and selection criteria
that \citet{steidel1999} use but to a deeper magniude
in the Keck Deep Fields which cover a total area of 169 arcmin$^2$
and consist of five fields.
\citet{iwata2003} found LBGs at $z\sim 5$ using \vic, \icz\ colors in
a 575 arcmin$^2$ area including the Hubble Deep Field North.
\citet{capak2004} obtained number counts of LBGs at $z\sim 4$
selected by \br, \ric\ colors and LBGs at $z\sim 5$ selected by \vic,
\icz\ colors from a 720 arcmin$^2$ area centered on the Hubble Deep
Field North.
Our limiting magnitudes are among the faintest ones
in the LBG samples for $z\sim 4$ and 5 constructed to date.

\begin{deluxetable}{cccccclcl}
  \rotate
  \tablecolumns{9}
  \tablewidth{0pt}
  \tablecaption{LBG surveys in the literature\label{tbl:ref}}
  \tablehead{
    \colhead{Sample}
    & \colhead{Area [arcmin$^2$]}
    & \multicolumn{3}{c}{$m_\mathrm{limit}$}
    & \colhead{\hspace{0.5cm} $N$\hspace{0.5cm} }
    & \colhead{Selection}
    & \colhead{Fig. in this paper}
    & \colhead{Ref.}
  }
  \startdata 
$z\sim 4$ & \phn875 & $i'$ & $=$ & $26.85$ & $3,808$
  & \hspace{0.5cm} \br, \ri
  & Fig. \ref{fig:num-bri}, \ref{fig:lf-z4}, \ref{fig:lf}
  & \hspace{0.3cm} This study \\
$z\sim 5$ & \phn875 & $z'$ & $=$ & $26.05$ & \phn\phd539
  & \hspace{0.5cm} \vi, \iz
  & Fig. \ref{fig:num-viz}, \ref{fig:lf-z5}, \ref{fig:lf}
  & \hspace{0.3cm} This study \\
$z\sim 5$ & \phn875 & $z'$ & $=$ & $26.05$ & \phn\phd240
  & \hspace{0.5cm} \ri, \iz
  & Fig. \ref{fig:num-riz}, \ref{fig:lf-z5}, \ref{fig:lf}
  & \hspace{0.3cm} This study \\
\\
\\
$z\sim 3$ & 1046 & $\mathfrak{R}$ & $=$ & $25.5\phn$ & $1,270$
  & \hspace{0.5cm} \ug, \gr
  & Fig. \ref{fig:lf}\phd\ \phn\phn\phd\ \phn\phn
  & \hspace{0.3cm} \citet{steidel1999} \\
$z\sim 4$ & \phn828 & $I$ & $=$ & $25.0\phn$ & \phn\phn207
  & \hspace{0.5cm} \gr, \rrii
  & Fig. \ref{fig:num-bri}, \ref{fig:lf-z4}\phd\ \phn\phn
  & \hspace{0.3cm} \citet{steidel1999} \\
$z\sim 4$ & \phn543 & $i'$ & $=$ & $26.3\phn$ & $1,438$
  & \hspace{0.5cm} \br, \ri
  & Fig. \ref{fig:num-bri}, \ref{fig:lf-z4}\phd\ \phn\phn
  & \hspace{0.3cm} \citet{ouchi2004} \\
$z\sim 4$ & \phn720 & $I_c$ & $=$ & $25.5\phn$ & ?
  & \hspace{0.5cm} \br, \ric
  & Fig. \ref{fig:num-bri}\phd\ \phn\phn\phd\ \phn\phn
  & \hspace{0.3cm} \citet{capak2004} \\
$z\sim 4$ & \phn169 & $\mathfrak{R}$ & $=$ & $27.0\phn$ & \phn\phd427
  & \hspace{0.5cm} \gr, \rrii
  & Fig. \ref{fig:lf-z4}\phd\ \phn\phn\phd\ \phn\phn
  & \hspace{0.3cm} \citet{sawicki2006} \\
$z\sim 5$ & \phn575 & $I_c$ & $=$ & $25.95$ & \phn\phd305
  & \hspace{0.5cm} \vic, \icz
  & Fig. \ref{fig:num-viz}, \ref{fig:lf-z5}\phd\ \phn\phn
  & \hspace{0.3cm} \citet{iwata2003} \\
$z\sim 5$ & \phn616 & $z'$ & $=$ & $25.8\phn$ & \phn\phd246
  & \hspace{0.5cm} \vi, \iz
  & Fig. \ref{fig:num-viz}, \ref{fig:lf-z5}\phd\ \phn\phn
  & \hspace{0.3cm} \citet{ouchi2004} \\
$z\sim 5$ & \phn720 & $I_c$ & $=$ & $25.5\phn$ & ?
  & \hspace{0.5cm} \vic, \icz
  & Fig. \ref{fig:num-viz}\phd\ \phn\phn\phd\ \phn\phn
  & \hspace{0.3cm} \citet{capak2004} \\
$z\sim 5$ & \phn616 & $z'$ & $=$ & $25.8\phn$ & \phn\phn\phd68
  & \hspace{0.5cm} \ri, \iz
  & Fig. \ref{fig:num-riz}, \ref{fig:lf-z5}\phd\ \phn\phn
  & \hspace{0.3cm} \citet{ouchi2004} \\
$z\sim 6$ & \phn\phn12\tablenotemark{\dagger} & $z'$ & $=$ & $29.5\phn$
  & \phn\phd506
  & \hspace{0.5cm} \iizz, \Vz
  & Fig. \ref{fig:lf}\phd\ \phn\phn\phd\ \phn\phn
  & \hspace{0.3cm} \citet{bouwens2005} \\
$z\sim 6$ & \phn767 & $z_\mathrm{R}$ & $=$ & $25.4\phn$ & \phn\phn\phd12
  & \hspace{0.5cm} \izr, \zbzr
  & Fig. \ref{fig:lf}\phd\ \phn\phn\phd\ \phn\phn
  & \hspace{0.3cm} \citet{shimasaku2005} \\
  \enddata 
  \tablenotetext{\dagger\ }{%
\citet{bouwens2005} incorporated the faint data
from the Hubble Ultra Deep Field (12 arcmin$^2$) with
two other data sets of shallower depths:
the two GOODS fields ($\sim$ 170 arcmin$^2$ each)
and the two Ultra Deep Field-Parallel fields (20 arcmin$^2$ in total).
}
\end{deluxetable} 

In general, number counts measurements of LBGs are dependent
on detection completeness, photometric systems used, and selection
criteria.
It is found that our counts match very well with those of
\citet{ouchi2004} for all the samples except for brightest bins.
While we use MAG\_AUTO in SExtractor as total magnitudes of objects,
\citet{ouchi2004} calculate total magnitudes from
aperture magnitudes through a fixed aperture correction of $-0.2$ mag.
The correction value, however, actually can range to as much as $-0.4$
mag for bright LBGs.
A shift of the magnitudes of \citet{ouchi2004} by $-0.2$ mag
($=-0.4-(-0.2)$)
can settle the disagreement between Ouchi et al.'s (2004)
and our counts.
The counts of \citet{capak2004} are lower
than our counts for $z\sim 4$ widely at faint magnitudes
and for $z\sim 5$ at the faintest magnitude of \citet{capak2004}.
The lower detection completeness of their data might be a major
cause of this inconsistency;
the $5\sigma$ limiting magnitude of their data is about 1 mag
brighter than that of the SDF Project data.
For $z\sim 4$ LBGs, our measurements agree with those of
\citet{steidel1999},
although Steidel et al.'s (1999) measurements are limited to objects
brighter than 25 mag.
At faint magnitudes, however, our measurements are higher
than those of \citet{sawicki2006}.
For $z\sim 5$ LBGs, over a wide range of bright magnitudes ($m<25$),
a discrepancy is found between the counts
of \citet{iwata2003} and either ours or those of \citet{capak2004}
who surveyed almost the same field as \citet{iwata2003};
Iwata et al.'s (2003) counts are higher about a factor of 3 -- 10.
The selection criteria of \citet{iwata2003} are
somewhat looser than ours. 
Therefore, as \citet{ouchi2004} and \citet{capak2004} claim,
the LBG sample of
\citet{iwata2003} could be strongly contaminated by foreground galaxies.

\subsection{Completeness} \label{subsec:lbg-comp}

The completeness at a given apparent magnitude and redshift is
defined as the ratio of LBGs which are detected and also
pass the selection criteria, to all the LBGs with the
given apparent magnitude and redshift actually present in the universe.
In general, completeness decreases as apparent magnitude goes
fainter and as redshift goes away from the central redshift of
the sample.
The completeness of each LBG sample is estimated as functions of
apparent magnitude and redshift through a Monte Carlo simulation.

\begin{figure}[h]
  \begin{center}
    \includegraphics[scale=0.5]{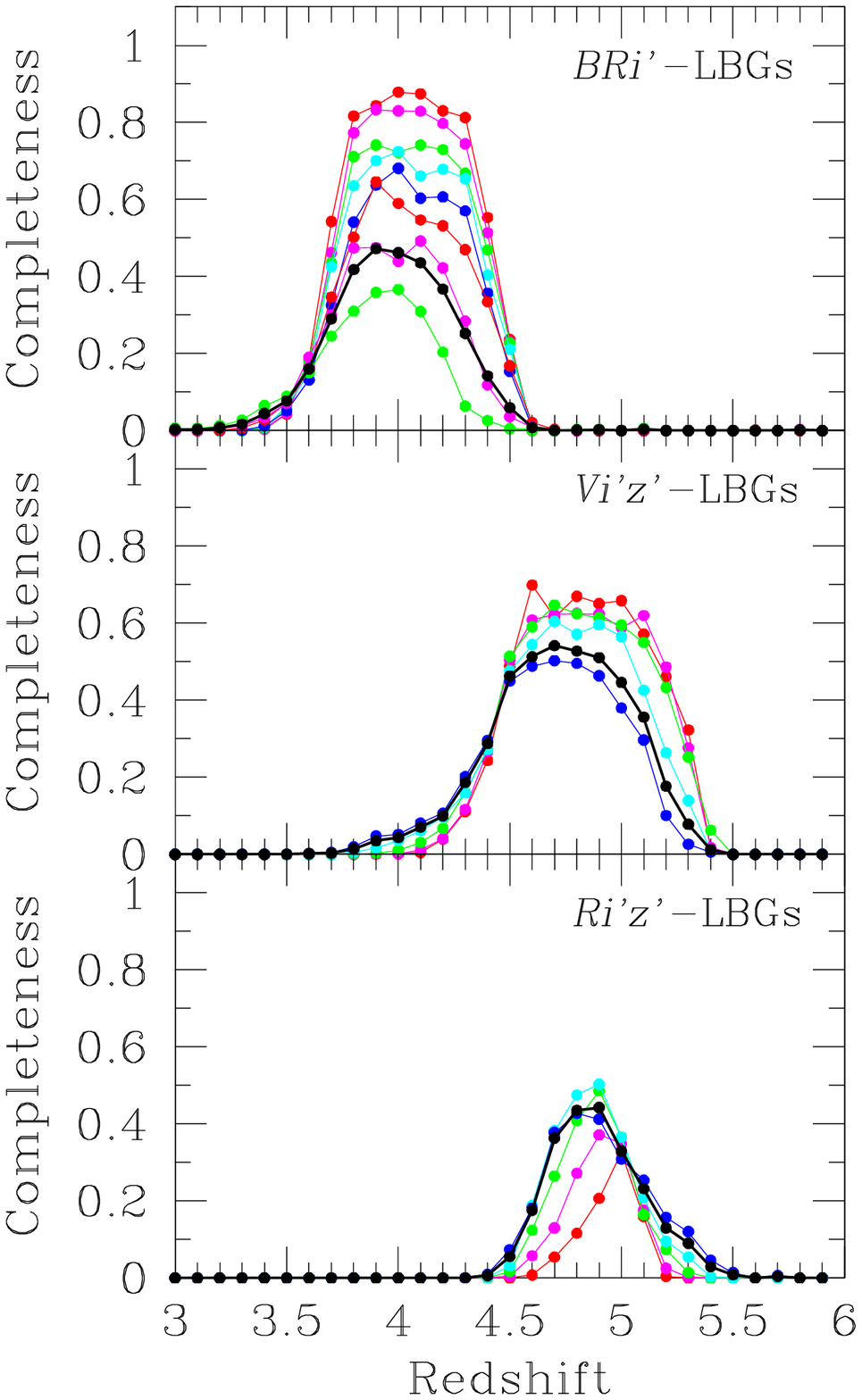}
  \end{center}
  \caption{%
Completeness against the redshift of objects with different apparent
magnitude
for our LBG samples.
In the panel of $BRi'$-LBGs,
the red, magenta, green, cyan, blue, red, magenta, and green
(from top to bottom) lines denote the completeness for
$i'=23.45$, 23.95, 24.45, 24.95, 25.45, 25.95, 26.45, and 26.95,
respectively.
In the panels of $Vi'z'$-LBGs and $Ri'z'$-LBGs,
the red, magenta, green, cyan, and blue lines denote the completeness
for $z'=24.15$, 24.65, 25.15, 25.65, and 26.15,
respectively.
The thick black lines in all panels
indicate the magnitude-weighted completeness.
\label{fig:comp}}
\end{figure}

In the simulation, we generate artificial LBGs over an apparent
magnitude range of $23.1\le m_{i'}\le 26.6$ for $BRi'$-LBGs and
$23.8\le m_{z'}\le25.8$ for $Vi'z'$-LBGs and $Ri'z'$-LBGs with an
interval of $\Delta m=0.5$, and over a redshift range of
$z=3.0\le z\le 5.9$ with an interval of $\Delta z = 0.1$.
All LBGs are assumed to have the same intrinsic spectrum described
in \S\ref{subsec:lbg-select}, but with values
of $E(B-V)=0.0$, $0.1$, $0.2$, $0.3$, and $0.4$.
Thus, $1,200$ ($8\times 30\times 5$) objects for $BRi'$-LBGs
and $750$ ($5\times 30\times 5$) objects
for $Vi'z'$-LBGs and $Ri'z'$-LBGs
are generated.
We assume that the shape of LBGs is PSF-like
and assign apparent sizes to the artificial LBGs so that their size
distribution measured by SExtractor
(the peak FWHM is $\approx 1.''00 - 1.''04$)
matches that of the observed LBG candidates.
The artificial LBGs are then distributed randomly on the original
images after adding Poisson noise according to their magnitudes, and
object detection and photometry are performed in the same manner
as done for real objects.
A sequence of these processes is repeated 100 times to obtain
statistically accurate values of completeness.
In the simulation, the completeness for a given apparent magnitude,
redshift, and $E(B-V)$ value
can be defined as the ratio in number of the simulated LBGs
which are detected and also satisfy the selection criteria, to all
the simulated objects with the given magnitude, redshift,
and $E(B-V)$ value.
We calculate the completeness of the LBG samples
by taking a weighted average of the completeness
for each of the five $E(B-V)$ values.
The weight is taken using the $E(B-V)$ distribution function of
$z\sim 4$ LBGs derived by
\citet{ouchi2004}(open histogram in the bottom panel of their Fig.20),
which is corrected for
incompleteness due to selection biases.
We plot the recovered artificial objects in the two-color diagrams
by taking the number of objects with each $E(B-V)$ value
according to the adopted $E(B-V)$ distribution function.
We verify that the distribution well resemble that of real objects.
Hence, our modeling is reasonably realistic.
The resulting completeness, $p(m,z)$, is shown in \reffig{comp}.

In order to examine the uniformity of completeness over our images,
we split the images into two regions
(a center region and a periphery region), each with the same area,
and recalculate the completeness for each region.
The non-uniformity is found to be small;
the variation between the center region and the periphery region
is within 10 \%.

For each of the three LBG samples,
magnitude-weighted redshift distribution
functions are derived from $p(m,z)$ (thick black lines in
\reffig{comp}) by averaging the magnitude-dependent completeness
weighted by the number of LBGs in each magnitude bin.
The average redshift, $\bar{z}$, and its standard deviation, $s_z$,
are calculated to be $\bar{z}=4.0$ and $s_z=0.3$ for $BRi'$-LBGs,
$\bar{z}=4.7$ and $s_z=0.3$ for $Vi'z'$-LBGs, and $\bar{z}=4.9$ and
$s_z=0.2$ for $Ri'z'$-LBGs.
We also compute magnitude-weighted redshift distribution functions
where the weighting is done by the number of
spectroscopically identified LBGs (dotted lines in \reffig{spec-nz}),
to obtain the average redshift
of $\bar{z}_\mathrm{spec}=4.0$ for $BRi'$-LBGs,
$\bar{z}_\mathrm{spec}=4.8$ for $Vi'z'$-LBGs,
and $\bar{z}_\mathrm{spec}=4.9$ for $Ri'z'$-LBGs.
These distribution functions are consistent with the
redshift distributions of spectroscopic samples
shown in \reffig{spec-nz}.

Although the above modeling well reproduces
the observed distribution of objects in the two-color diagrams
and the observed redshift distribution functions,
we further explore to what extent the results would be affected
by adopting different models.
To begin with, we recalculate the completeness using model spectra
with two other ages, 0.01 Gyr and 0.5 Gyr, to find that
the completeness changes only a few percent.
Then, we examine two extreme cases of dust extinction
that all model spectra have $E(B-V)=0.0$ and 0.4.
Changes in completeness are found to be negligibly small
in either case except for $Ri'z'$-LBGs.
Next, we examine the effect of changing the absorption by
the intergalactic medium.
If we shift the amount of attenuation to $\pm 1\sigma$
from the average amount given in \citet{madau1995},
we obtain quite different completeness values.
However, the redshift distribution functions derived are
unrealistic, because they are strongly inconsistent with the
distributions of spectroscopic objects.
We also adopt \citet{meiksin2006}'s prescription for absorption
by the intergalactic medium,
to find completeness values similar to those based on
\citet{madau1995}'s.

\subsection{Contamination by Interlopers} \label{subsec:lbg-contam}

We estimate the fraction of low-redshift interlopers in the LBG
samples by a Monte Carlo simulation as follows.
For the boundary redshift, $z_0$, between interlopers and LBGs,
$z_0=3.5$ is adopted for $BRi'$-LBGs, $z_0=4.0$ for $Vi'z'$-LBGs,
and $z_0=4.5$ for $Ri'z'$-LBGs.

\begin{figure}[h]
  \begin{center}
    \includegraphics[scale=0.5]{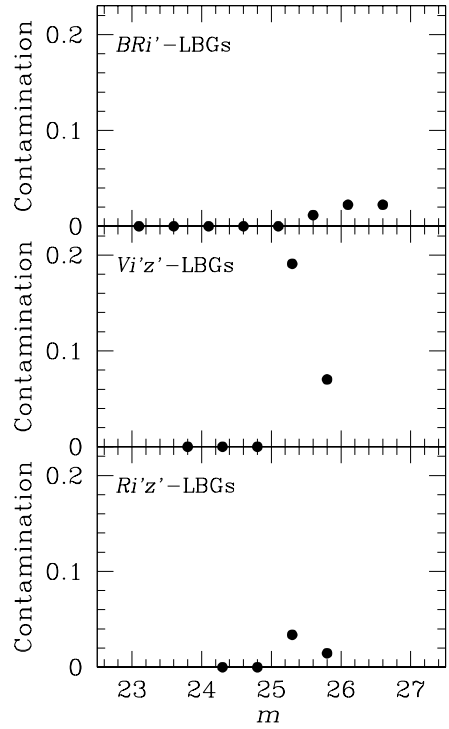}
  \end{center}
  \caption{%
Fraction of interlopers as a function of magnitude for our LBG samples.
\label{fig:contam}}
\end{figure}

We use objects in the Hubble Deep Field North (HDFN), for which
best-fit spectra and photometric redshifts are given by
\citet{furusawa2000}, as a template of the color, magnitude, and
redshift distribution of foreground galaxies, and generate 929
artificial objects which mimic the HDFN objects.
The apparent sizes of the artificial objects are adjusted so that
the size distribution recovered from the simulation is similar to
that of the real objects in our catalogs.
We distribute the artificial objects randomly on the original images
after adding Poisson noise according to their magnitudes, and
perform object detection and photometry in the same manner as
employed for real objects.
A sequence of these processes is repeated 100 times.
In the simulation, the number of interlopers can be defined as the
number of the simulated objects with low redshift ($z<z_0$) which are
detected and also satisfy the selection criteria for LBGs.
The number of interlopers expected in an LBG sample can then be
calculated by multiplying the raw number by a scaling factor
which corresponds to the
ratio of the area of the SDF (875 arcmin$^2$) to
the area of the HDFN
multiplied by the repeated times (100 $\times $ 3.92 arcmin$^2$).
\reffig{contam} shows the fraction
of interlopers for our LBG samples as a function of magnitude.
For the $BRi'$-LBG and $Ri'z'$-LBG samples, the fraction is found to be
less than 5\% at any magnitude.
For the $Vi'z'$-LBG sample, the contamination is
higher but at most $\simeq 20$\%.
Most of the interlopers are in the redshift range of $0.2\le z\le 0.8$,
as predicted from Figures \ref{fig:model-bri} -- \ref{fig:model-riz}.
The rest are objects at redshifts which are close to the
boundary redshift of each sample.
The fraction of interlopers only at low redshifts (i.e., not
at near the boundary redshifts) for each whole sample is
around 1\% for the $BRi'$-LBG and $Ri'z'$-LBG samples,
and around 9\% for the $Vi'z'$-LBG sample.

The number density and the redshift distribution of galaxies 
in the HDFN may be largely different from the cosmic averages 
because the HDFN is a very small field.
However, the contamination by interlopers for each of 
the three LBG samples calculated above is very low. 
We therefore expect that the uncertainty in contamination 
due to a possible (large) cosmic variance in the HDFN galaxies 
will not be a significant source of the error in the luminosity 
functions of LBGs derived in the next section.

\clearpage

\section{LUMINOSITY FUNCTIONS AT REST-FRAME ULTRAVIOLET WAVELENGTHS}
\label{sec:lf}

We derive the luminosity functions (LFs) of LBGs at $z\sim 4$ -- 5 by
applying the ``effective volume'' method \citep{steidel1999}.
The Monte Carlo simulation in \S\ref{subsec:lbg-comp} yields the
effective survey volume as a function of apparent magnitude,
\begin{eqnarray}
  V_\mathrm{eff}(m) &=& \int_{z_0}^\infty p(m,z)\frac{dV(z)}{dz}\,dz,
\end{eqnarray}
where $p(m,z)$ is the probability that a galaxy of apparent magnitude
$m$ at redshift $z$ is detected and passes the selection
criteria (i.e., the completeness in \S\ref{subsec:lbg-comp} for
magnitude $m$ and redshift $z$), $\frac{dV(z)}{dz}$ is the
differential comoving volume at redshift $z$ for a solid angle of
the surveyed area (875 arcmin$^2$), and $z_0$ is the boundary redshift
between low-redshift galaxies and LBGs.
The effective volume obtained for each magnitude bin is provided
in Tables \ref{tbl:lf-bri} -- \ref{tbl:lf-riz}.

\begin{deluxetable}{cccc} 
  \tablecolumns{4}
  \tablewidth{0pt}
  \tablecaption{%
Number of LBGs detected, number of interlopers,
and effective survey volume for $BRi'$-LBGs.
\label{tbl:lf-bri}}
  \tablehead{
    \colhead{\hspace{1cm} Magnitude Range ($i'$)\hspace{1cm} }
    & \colhead{\hspace{0.5cm} $N_\mathrm{raw}$\tablenotemark{a}\hspace{0.5cm} }
    & \colhead{\hspace{0.5cm} $N_\mathrm{interloper}$\tablenotemark{b}\hspace{0.5cm} }
    & \colhead{\hspace{0.5cm} $V_\mathrm{eff}$\tablenotemark{c}\hspace{0.5cm} }
  }
  \startdata 
22.85 -- 23.35\dotfill & \phn\phn\phn 3 & \phn 0\phd\phn & $1.74\times 10^6$ \\
23.35 -- 23.85\dotfill & \phn\phn 12    & \phn 0\phd\phn & $1.63\times 10^6$ \\
23.85 -- 24.35\dotfill & \phn\phn 68    & \phn 0\phd\phn & $1.49\times 10^6$ \\
24.35 -- 24.85\dotfill & \phn 231       & \phn 0\phd\phn & $1.39\times 10^6$ \\
24.85 -- 25.35\dotfill & \phn 447       & \phn 0\phd\phn & $1.22\times 10^6$ \\
25.35 -- 25.85\dotfill & \phn 763       & \phn 8.9       & $1.15\times 10^6$ \\
25.85 -- 26.35\dotfill & 1093           & 24.6           & $8.77\times 10^5$ \\
26.35 -- 26.85\dotfill & 1191           & 26.8           & $5.70\times 10^5$ \\
  \enddata 
  \tablenotetext{a\ }{Number of LBGs detected.}
  \tablenotetext{b\ }{%
Number of interlopers estimated from our simulations
(see \S\ref{subsec:lbg-contam}).}
  \tablenotetext{c\ }{%
Effective survey volume for 875 arcmin$^2$
in units of Mpc$^3$.}
\end{deluxetable} 

\begin{deluxetable}{cccc} 
  \tablecolumns{4}
  \tablewidth{0pt}
  \tablecaption{%
Number of LBGs detected, number of interlopers,
and effective survey volume for $Vi'z'$-LBGs.
\label{tbl:lf-viz}}
  \tablehead{
    \colhead{\hspace{1cm} Magnitude Range ($z'$)\hspace{1cm} }
    & \colhead{\hspace{0.5cm} $N_\mathrm{raw}$\tablenotemark{a}\hspace{0.5cm} }
    & \colhead{\hspace{0.5cm} $N_\mathrm{interloper}$\tablenotemark{b}\hspace{0.5cm} }
    & \colhead{\hspace{0.5cm} $V_\mathrm{eff}$\tablenotemark{c}\hspace{0.5cm} }
  }
  \startdata 
23.55 -- 24.05\dotfill & \phn\phn 1 & \phn 0\phd\phn & $1.34\times 10^6$ \\
24.05 -- 24.55\dotfill & \phn 12    & \phn 0\phd\phn & $1.31\times 10^6$ \\
24.55 -- 25.05\dotfill & \phn 56    & \phn 0\phd\phn & $1.31\times 10^6$ \\
25.05 -- 25.55\dotfill & 152        & 29.0           & $1.17\times 10^6$ \\
25.55 -- 26.05\dotfill & 318        & 22.3           & $9.66\times 10^5$ \\
  \enddata 
  \tablenotetext{a\ }{Number of LBGs detected.}
  \tablenotetext{b\ }{%
Number of interlopers estimated from our simulations
(see \S\ref{subsec:lbg-contam}).}
  \tablenotetext{c\ }{%
Effective survey volume for 875 arcmin$^2$
in units of Mpc$^3$.}
\end{deluxetable} 

\begin{deluxetable}{cccc} 
  \tablecolumns{4}
  \tablewidth{0pt}
  \tablecaption{%
Number of LBGs detected, number of interlopers,
and effective survey volume for $Ri'z'$-LBGs.
\label{tbl:lf-riz}}
  \tablehead{
    \colhead{\hspace{1cm} Magnitude Range ($z'$)\hspace{1cm} }
    & \colhead{\hspace{0.5cm} $N_\mathrm{raw}$\tablenotemark{a}\hspace{0.5cm} }
    & \colhead{\hspace{0.5cm} $N_\mathrm{interloper}$\tablenotemark{b}\hspace{0.5cm} }
    & \colhead{\hspace{0.5cm} $V_\mathrm{eff}$\tablenotemark{c}\hspace{0.5cm} }
  }
  \startdata 
24.05 -- 24.55\dotfill & \phn\phn 3 & 0\phd\phn & $3.32\times 10^5$ \\
24.55 -- 25.05\dotfill & \phn 18    & 0\phd\phn & $4.60\times 10^5$ \\
25.05 -- 25.55\dotfill & \phn 66    & 2.2       & $5.54\times 10^5$ \\
25.55 -- 26.05\dotfill & 153        & 2.2       & $5.69\times 10^5$ \\
  \enddata 
  \tablenotetext{a\ }{Number of LBGs detected.}
  \tablenotetext{b\ }{%
Number of interlopers estimated from our simulations
(see \S\ref{subsec:lbg-contam}).}
  \tablenotetext{c\ }{%
Effective survey volume for 875 arcmin$^2$
in units of Mpc$^3$.}
\end{deluxetable} 

The number density of LBGs corrected for incompleteness
and contamination is computed as:
\begin{eqnarray}
  \phi (m) &=&
  \frac{N_\mathrm{raw}(m) - N_\mathrm{interloper}(m)}{V_\mathrm{eff}(m)},
\end{eqnarray}
where $N_\mathrm{raw}(m)$ and $N_\mathrm{interloper}(m)$
are the number of LBGs detected and the number of interlopers
which is estimated from the simulations, respectively,
in an apparent magnitude bin of $m$.
Tables \ref{tbl:lf-bri} -- \ref{tbl:lf-riz} also give
$N_\mathrm{raw}(m)$ and $N_\mathrm{interloper}(m)$.

The number densities of LBGs in faint bins, especially 
the faintest bins, could be largely overestimated owing to 
Eddington bias from objects in fainter bins and/or
beyond the limiting magnitude 
of the sample which have very large photometric errors 
and probably have a larger number density.
We evaluate the effect of this bias for each LBG sample as follows.
First, we construct a deeper sample of LBGs down to a magnitude 
limit fainter by 0.5 mag than the original value 
(for example, the new limit is $i' \le 27.35$ for $BRi'$-LBGs).
Then, using this deeper sample, we carry out Monte Carlo 
simulations similar to those made in \S 3.3, 
and derive $V_{\rm eff}(m)$ to calculate $\phi(m)$ to a deeper
magnitude.
From the simulations, we compute for each magnitude bin
the fractions of LBGs
which scatter out to other magnitude bins.
Finally, we recalculate $\phi(m)$ by correcting the original
number densities for the scatter along the magnitude axis.
The number density of LBGs derived in this way is found to be 
consistent with the original calculation within statistical errors, 
down to the original limiting magnitude. 
This means that Eddington bias is negligibly small for our samples.

The LF in the rest-frame ultraviolet ($\simeq 1500$ \AA) absolute
magnitude, $M_\mathrm{UV}$, is obtained by
converting $\phi(m)$ into $\phi(M_\mathrm{UV}(m))$.
For each LBG sample, we calculate the absolute magnitude
of LBGs from their apparent magnitude using the
average redshift of the sample
by
\begin{eqnarray}
  M_\mathrm{UV}
  &=& m
      + 2.5\log (1+\bar{z})
      - 5\log \left(d_L(\bar{z})\right) + 5 \nonumber\\
  &&  {} + (m_\mathrm{UV}-m),
\label{eq:mag}
\end{eqnarray}
where $d_L$ is the luminosity distance in units of pc.
For apparent magnitude, $m$, we use $i'$ magnitude for $BRi'$-LBGs,
and $z'$ magnitude for $Vi'z'$-LBGs and $Ri'z'$-LBGs.
The last term of Eq.\ref{eq:mag}, $(m_\mathrm{UV}-m)$, is
the difference between the magnitude at rest-frame
1500A and the magnitude in the bandpass being considered
in the rest-frame.
Using the representative galaxy models given in
\S\ref{subsec:lbg-select},
we find $(m_\mathrm{UV}-m)$ to be negligible, or no larger than 
$0.03$ mag, over the redshift ranges selected by the color selection
criteria we adopt.
Thus we set $(m_\mathrm{UV}-m)$ to zero in Eq. \ref{eq:mag}.
We assume that all of the LBGs have the average redshift of the sample.
Indeed, for each of the three LBG samples, 
varying the redshift of an object over the standard 
deviation of the redshift distribution of the sample 
changes its absolute magnitude by not more than $0.1$ mag.

\begin{figure}[t]
  \begin{center}
    \includegraphics[scale=0.4]{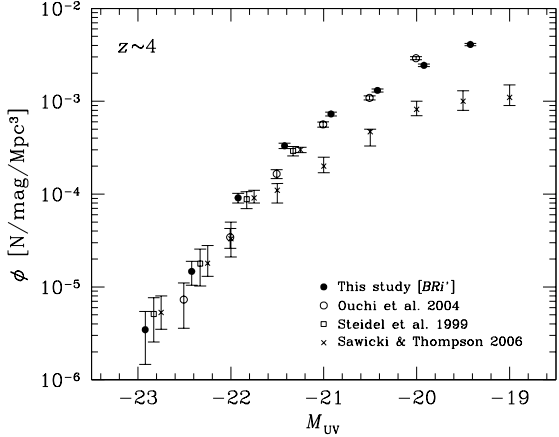}
  \end{center}
  \caption{%
UV luminosity functions (LFs) of LBGs at $z\sim 4$.
Our data are shown by the filled circles.
The open circles, open squares, and crosses are
from \citet{ouchi2004}, \citet{steidel1999}, and \citet{sawicki2006},
respectively.
The error bars for our data and Ouchi et al.'s (2004)
reflect Poisson errors, and those for the other two data include
both Poisson errors and an estimate of field-to-field variance from
their multiple fields.
\label{fig:lf-z4}}
\end{figure}

\begin{figure}[t]
  \begin{center}
    \includegraphics[scale=0.4]{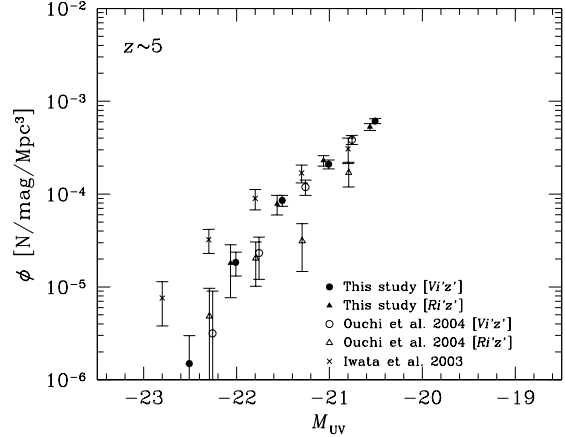}
  \end{center}
  \caption{%
UV luminosity functions (LFs) of LBGs at $z\sim 5$.
Our data from $Vi'z'$-LBG sample and $Ri'z'$-LBG sample are shown
by the filled circles and filled triangles, respectively.
The open circles and open triangles are from
Ouchi et al.'s (2004) $Vi'z'$-LBG sample and $Ri'z'$-LBG sample,
respectively.
The crosses are from \citet{iwata2003}.
The error bars reflect Poisson errors.
\label{fig:lf-z5}}
\end{figure}

In Figures \ref{fig:lf-z4} and \ref{fig:lf-z5}, the LFs
at $z\sim 4$ and $z\sim 5$ are plotted respectively,
along with those of other authors (See \reftbl{ref}.).
Note that \citet{capak2004} do not derive the LFs.
Our LFs are in excellent agreement
with those of \citet{ouchi2004}
up to their faintest magnitudes for both redshifts.
This agreement appears reasonable, since \citet{ouchi2004}
and this study are based on the same field (the SDF)
and adopt almost the same selection criteria.
Our LFs reach fainter magnitudes than theirs.
However, there is a discrepancy between our $z\sim 4$ LF
and that of \citet{sawicki2006} in the Keck Deep Fields
at the faint end;
the faint-end slope of our LF is steeper
than that of \citet{sawicki2006}.
\citet{gabasch2004} also
derived the galaxy LFs at rest-frame ultraviolet wavelengths
in the FORS Deep Field using photometric redshifts
and obtained a similar result to that of \citet{sawicki2006}.
The reason for this inconsistency is not clear.
We note, however, that our raw counts before the correction of
incompleteness already exceed the
Sawicki \& Thompson's (2006) corrected
counts, and thus that our higher values are not due to
overcorrections.
Additionally, our data have a sky coverage five times larger than
that of the total of the five Keck Deep Fields.
Thus, our LF is more robust against a possible cosmic variance
on large scales,
although the five separate Keck Deep Fields are less affected by
cosmic variance than a single continuous field
of the same sky coverage.
At $z\sim 5$, our LF is different from that of \citet{iwata2003}
at bright magnitudes.
One possible explanation for this difference is that
\citet{iwata2003} may select a large number of contaminants
as mentioned in \S\ref{subsec:lbg-num},
resulting in the higher number density of bright LBGs.

We find a good agreement between our $z\sim 5$ LFs obtained from
the $Vi'z'$-LBG sample and the $Ri'z'$-LBG sample.
The redshift ranges of these two samples are almost the same.
This consistency in the results from the two independent manners
provides strong support for LFs we obtained.
As the $Vi'z'$-LBG sample contains a statistically larger number
of objects and has higher completeness,
we only use results from the $Vi'z'$-LBG sample in the following
analysis for $z\sim 5$.
These LFs obtained at $z\sim 4$ and 5 are found to be well
reproduced by a semi-analytic model combined with high-resolution
N-body simulations \citep{nagashima2005} when
the observational selection effects are taken into account.
We should, however, point out that the Lyman-break technique 
cannot select all star-forming galaxies at targeted redshifts.
On the basis of spectroscopy of galaxies in an optically-selected, 
flux-limited sample, Le Fevre et al. (2005) claim that a large 
fraction of galaxies at $z\sim 3$ -- 4 lie outside 
the selection boundaries for LBGs in two-color diagrams.
In addition, submm-selected galaxies \citep[e.g.,][]{smail1997}
and NIR-selected star-forming galaxies \citep[e.g.,][]{franx2003}
are often too faint in optical wavelengths  
to be selected as LBGs, because of strong absorption by dust.
Such dusty star-forming galaxies could contribute to 
the cosmic SFR density as much as do LBGs
\citep[e.g.,][]{chapman2005}.
Therefore, the far UV luminosity function and the star formation 
rate density derived in this paper are for galaxies with modest 
dust extinction and passing selection criteria for LBGs, 
and thus they may be greatly modified if all star-forming 
galaxies are included.

We fit the LFs with the Schechter function \citep{schechter1976}:
\begin{eqnarray}
  \lefteqn{\phi(L)dL}\hspace{0.1cm} \nonumber\\
     &=& \phi^\ast\left(\frac{L}{L^\ast}\right)^\alpha
         \exp\left(-\frac{L}{L^\ast}\right) d\left(\frac{L}{L^\ast}\right),
\end{eqnarray}
or expressed in terms of absolute magnitude:
\begin{eqnarray}
  \lefteqn{\phi(M_\mathrm{UV})dM_\mathrm{UV}}\hspace{0.1cm} \nonumber\\
  &=&
  \left(\frac{\ln 10}{2.5}\phi^\ast\right)
  \left(10^{-0.4(M_\mathrm{UV}-M_\mathrm{UV}^\ast)}\right)^{\alpha+1} \nonumber\\
  && \times\exp\left(-10^{-0.4(M_\mathrm{UV}-M_\mathrm{UV}^\ast)}\right)
dM_\mathrm{UV},
\end{eqnarray}
where  $\phi^\ast$, $L^\ast$ ($M_\mathrm{UV}^\ast$),
and $\alpha$ are parameters
to be determined from the data.
The parameter $\phi^\ast$ is a normalization factor which has a
dimension of number density, $L^\ast$ is a ``characteristic
luminosity'' (with an equivalent ``characteristic absolute
magnitude'', $M_\mathrm{UV}^\ast$),
and $\alpha$ gives the slope of the luminosity
function at the faint end.
For the $z\sim 5$ LF,
whose $\alpha$ has a rather large uncertainty,
we evaluate $\phi^\ast$ and $M_\mathrm{UV}^\ast$
with $\alpha$ fixed to the best-fit value for the $z\sim 4$ LF,
besides fitting with all three parameters left free.
The best-fit parameters obtained are listed in \reftbl{schechter}.

\begin{deluxetable}{cccccc} 
  \tablecolumns{6}
  \tablewidth{0pt}
  \tablecaption{Luminosity function parameters \label{tbl:schechter}}
  \tablehead{
    \colhead{\hspace{0.1cm} Sample Name\hspace{0.1cm} }
    & \colhead{\hspace{0.3cm} $\bar{z}$\hspace{0.3cm} }
    & \colhead{\hspace{0.3cm} $s_z$\hspace{0.3cm} }
    & \colhead{\hspace{0.3cm} $\phi^\ast$ [Mpc$^{-3}$]\hspace{0.3cm} }
    & \colhead{\hspace{0.3cm} $M_\mathrm{UV}^\ast$\hspace{0.3cm} }
    & \colhead{\hspace{0.3cm} $\alpha$\hspace{0.3cm} }
  }
  \startdata 
$BRi'$-LBGs  & 4.0 & 0.3 & $1.46^{+0.41}_{-0.35}\times 10^{-3}$
& $-21.14^{+0.14}_{-0.15}$ & $-1.82^{+0.09}_{-0.09}$ \\
\\
$Vi'z'$-LBGs & 4.7 & 0.3 & $0.58^{+1.04}_{-0.49}\times 10^{-3}$
& $-21.09^{+0.54}_{-0.74}$ & $-2.31^{+0.68}_{-0.60}$ \\
$Vi'z'$-LBGs & 4.7 & 0.3 & $1.23^{+0.44}_{-0.27}\times 10^{-3}$
& $-20.72^{+0.16}_{-0.14}$ & $-1.82$ (fixed) \\
  \enddata 
\end{deluxetable}

\section{%
EVOLUTION OF THE COSMIC STAR FORMATION ACTIVITY OVER $0\le z\le 6$}
\label{sec:dis}
\subsection{Evolution of the Luminosity Function}
\label{subsec:dis-lf}

\begin{figure}[t]
  \begin{center}
    \includegraphics[scale=0.4]{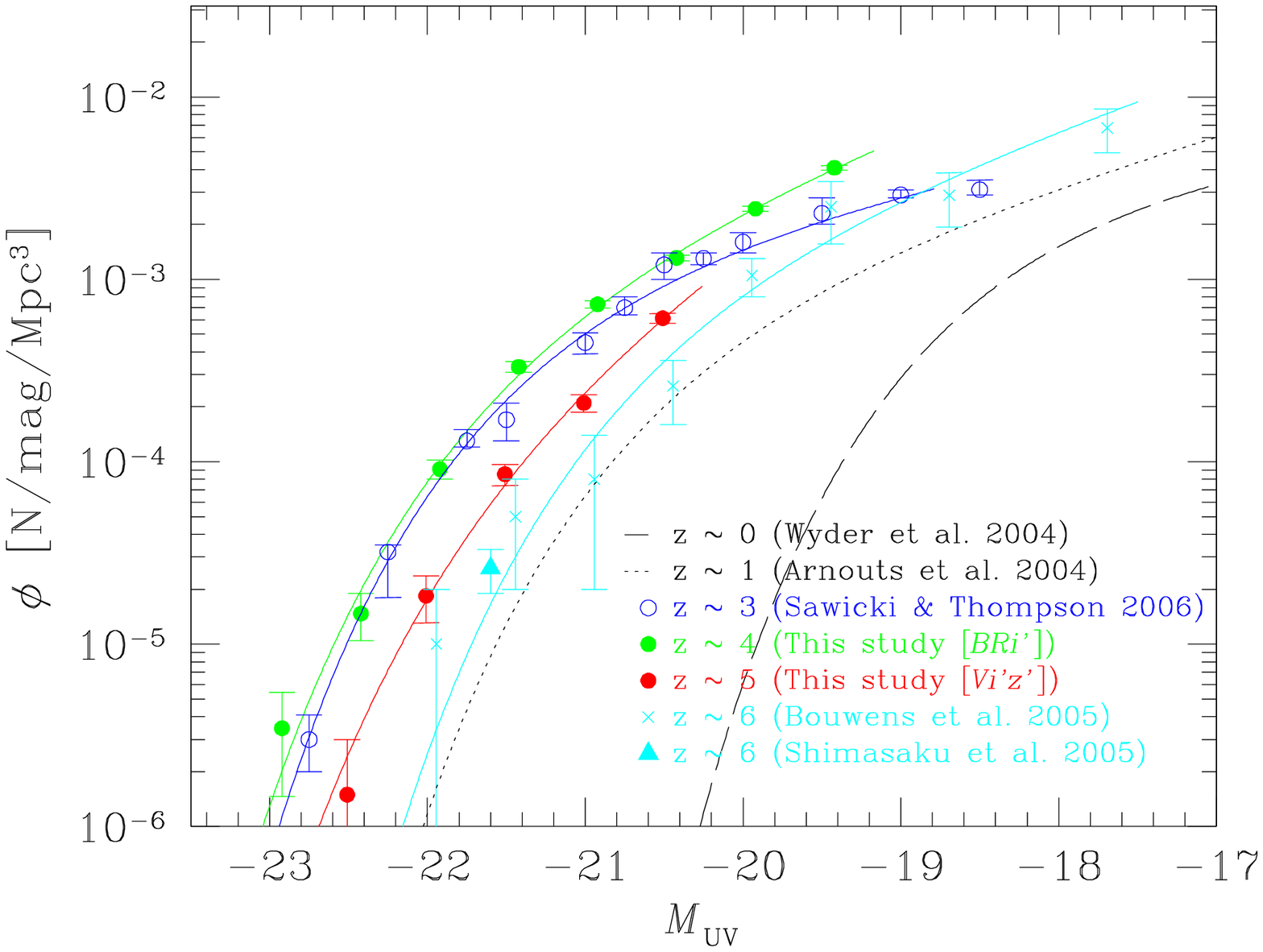}
  \end{center}
  \caption{%
UV luminosity functions (LFs) of LBGs at
$z\sim 3$ \citep[blue open circles:][]{sawicki2006},}
$z\sim 4$ (green filled circles: this study [$BRi'$-LBGs]),
$z\sim 5$ (red filled circles: this study [$Vi'z'$-LBGs]),
and $z\sim 6$ (cyan crosses: Bouwens et al. 2005;
cyan filled triangle: Shimasaku et al. 2005).
The error bars reflect Poisson errors.
The best-fit Schechter function is also shown with the solid line
for each data set except for that of \citet{shimasaku2005}.
The black dashed line and the black dotted line indicate
the LFs of UV-selected galaxies at $z\sim 0$ \citep{wyder2005},
and $z\sim 1$ \citep{arnouts2005}, respectively.
\label{fig:lf}
\end{figure}

We plot the LFs at $z\sim 4$ and 5 obtained in this study
with those of LBGs
at $z\sim 3$ \citep{sawicki2006}
and $z\sim 6$
(Bouwens et al. 2004a; Shimasaku et al. 2005 from observation of the
SDF in new filters sensitive to $\simeq 1\ \mu$m)
in \reffig{lf}.
The LFs of UV (1500\AA)-selected galaxies at $z\sim 0$
\citep{wyder2005} and
$z\sim 1$ \citep{arnouts2005} are presented as well for reference.
\reffig{lf} reveals clear evolution of the LF beyond $z\sim 4$.
From $z\sim 4$ to $z\sim 3$, we find no significant evolution
over the whole luminosity range.
\citet{sawicki2006} and \citet{gabasch2004} detect an increase in
faint galaxies across this redshift interval,
but such an increase is not seen in our data.
At lower redshifts, luminous galaxies strongly decrease
in number
from $z\gtrsim 2$ to $z\sim 0$,
as has been found by previous studies.

It is unlikely that the evolution in LF seen in \reffig{lf} is 
caused by the differences in selection criteria among 
the different samples.
We find, using spectra of model galaxies, 
that the criteria for LBGs at $z\sim 3$, 4, and 5 
select galaxies over similar ranges of age and $E(B-V)$.
Since $i'$-dropout galaxies at $z\sim 6$ are selected with
$i'-z'$ color alone,
they can have a wide range of $E(B-V)$. 
However, since no observation has reported a significant 
increase in dust extinction at $z>5$, it is reasonable to 
assume that the $i'$-dropout galaxies are a similar 
population to LBGs at lower redshifts.
Galaxies at $z\lesssim 1$ are selected in completely different ways 
from the Lyman-break technique. 
However, the differences in LF between $z\gtrsim 3$ and 
$z\lesssim 1$ seen in \reffig{lf} appear to be too large to be 
accounted for by possible selection effects. 
The dimming of LF seen for $z\lesssim 1$ will at least be 
qualitatively correct.

\begin{figure}[t]
  \begin{center}
    \includegraphics[scale=0.4]{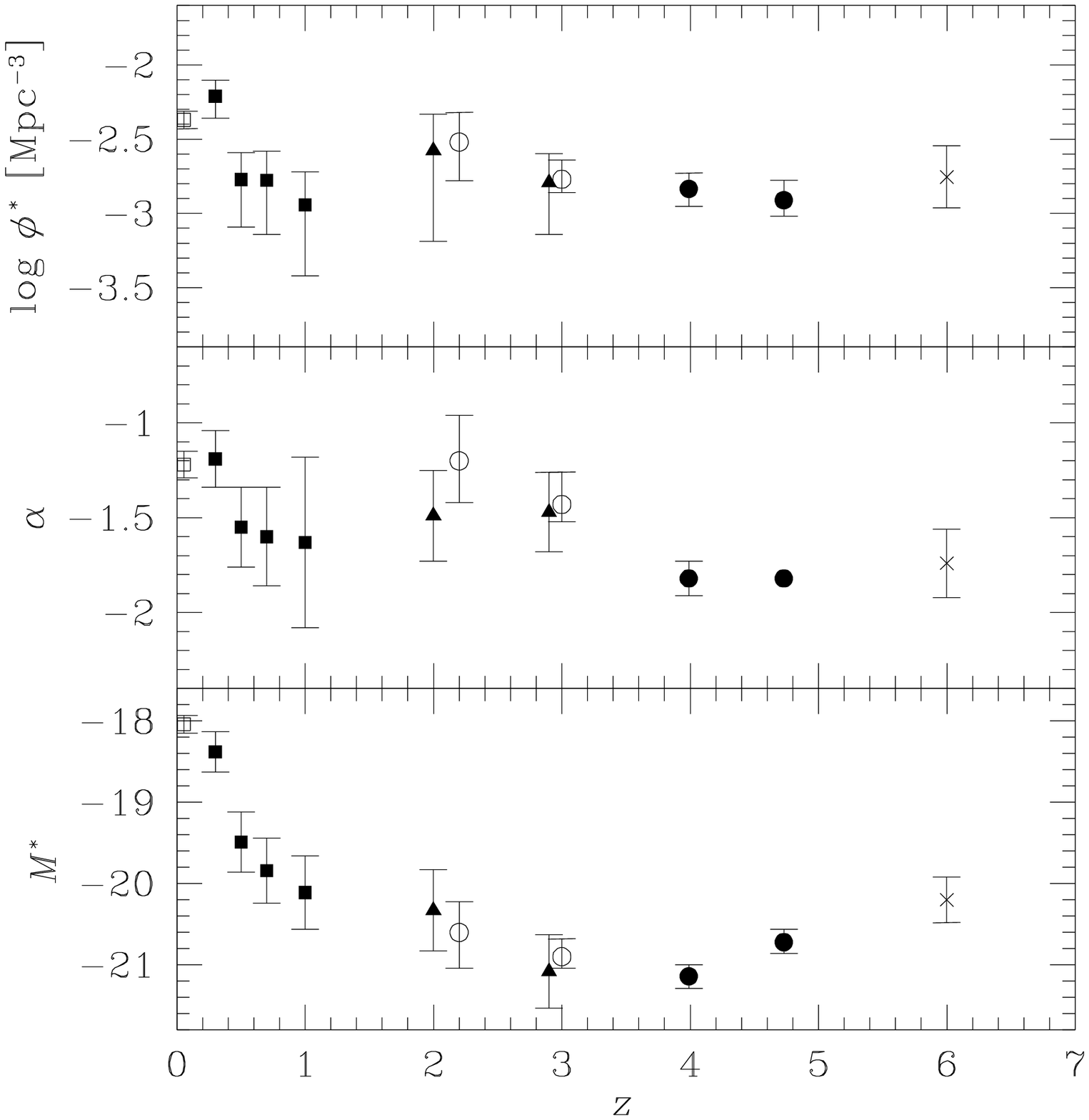}
  \end{center}
  \caption{%
Evolution of the luminosity function parameters with time.
The filled circles represent the measurements from this study.
The other symbols show measurements taken from the literature:
open squares from \citet{wyder2005},
filled squares from Arnouts et al. (2005; GALEX),
filled triangles from Arnouts et al. (2005; HDF),
open circles from \citet{sawicki2006},
and crosses from \citet{bouwens2005}.
For $z\sim 5$, we fixed $\alpha$ to the value
at $z\sim 4$ ($\alpha=-1.82$) in evaluating the other two parameters.
\label{fig:schechter}}
\end{figure}

\reffig{schechter} shows the evolution of Schechter parameters
over $0\le z\le 6$.
The filled circles represent our measurements.
The other symbols show measurements taken from the literature.
Note that the measurements
at $z\sim 2$ and $z\sim 3$ from \citet{arnouts2005}
may have large uncertainties,
because they are based solely on the HDFs which cover only a tiny area.
The normalization factor, $\phi^\ast$,
and the faint-end slope, $\alpha$,
appear to be almost constant over the redshift range of
$0.5\le z\lesssim 6$, followed by a significant change
at lower redshifts.
We note that $\alpha$ at $z\sim 5$ is fixed to the value
at $z\sim 4$ in evaluating the other two parameters
but the value at $z\sim 4$ is similar to that at $z\sim 6$.
From $z\sim 0.5$ to $z\sim 0$,
$\phi^\ast$ rises by a half order of magnitude and
$\alpha$ becomes shallower from $-1.6$ to $-1.2$.
On the other hand, the characteristic magnitude, $M^\ast$, strongly
and non-monotonically evolves over $0\le z\le 6$.
It brightens by about 1 mag from $z\sim 6$ to $z\sim 4$,
remains unchanged to $z\sim 3$, and then
a significant fading of about 3 mag occurs
from $z\sim 3$ to $z\sim 0$.
In other words, the evolution of the LF seen in \reffig{lf}
is accounted for by
a change in $M^\ast$ with cosmic time.
We point out, however, that there are different results about the LFs
at $z\sim 4$ and $z\sim 5$, as mentioned in \S \ref{sec:lf}.
Some groups claim that the faint-end slope is shallower
at higher redshifts
\citep{iwata2003,gabasch2004,sawicki2006}.
The characteristic magnitude of \citet{sawicki2006} at $z\sim 4$
is consistent with ours.
\citet{iwata2003} obtained an about 1 mag brighter value at $z\sim 5$
than ours.

\subsection{Cosmic Star Formation Rate Density}
\label{subsec:dis-sfr}

\subsubsection{%
Cosmic Star Formation Rate Density beyond $z\sim 3$}
\label{subsubsec:dis-sfr-lbg}

The ultraviolet luminosity density, $\rho_{L}$, at $z\sim 4$ and 5
is derived by integrating the LFs obtained in \S\ref{sec:lf} as:
\begin{eqnarray}
\rho_L &=& \int^\infty_{L_\mathrm{min}} L\,\phi(L)\,dL.
\end{eqnarray}
We consider two kinds of luminosity density.
One is the ``observed'' luminosity density, $\rho_L^\mathrm{obs}$,
which is derived from integration down to the faintest luminosity
of the sample, $L_\mathrm{min}=L_\mathrm{min}^\mathrm{obs}$.
This quantity, $\rho_L^\mathrm{obs}$, gives a lower limit of the
true luminosity density.
The faintest absolute magnitude of the sample at each redshift is 
$M_\mathrm{UV}=-19.2$ for $z\sim 4$ ($BRi'$-LBGs)
and $M_\mathrm{UV}=-20.3$ for $z\sim 5$ ($Vi'z'$-LBGs).
The other is the luminosity density derived by integrating
the LFs down to the same luminosity limit for all the measures.
This is important to properly account for possible evolution.
Following \citet{steidel1999},
we choose the limit to be $L_\mathrm{min}=0.1L^\ast_{z=3}$,
where
$L^\ast_{z=3}$ is the characteristic luminosity of the LF of $z\sim 3$
LBGs given by \citet{steidel1999}.
The luminosity $0.1L^\ast_{z=3}$ corresponds to $M_\mathrm{UV}=-18.6$.
We refer to this quantity as $\rho_L^{L>0.1L^\ast_{z=3}}$ hereafter.
When calculating
$\rho_L^{L>0.1L^\ast_{z=3}}$,
we extrapolate
the Schechter functions derived in \S\ref{sec:lf}
to fainter magnitudes
than the observation limits.
This extrapolation is modest for the $z\sim 4$ LBG sample ($BRi'$-LBGs),
since $L_\mathrm{min}^\mathrm{obs}$ of this sample
($M_\mathrm{UV}=-19.2$) is close to $0.1L^\ast_\mathrm{z=3}$,
On the other hand, the extrapolation is large for the $z\sim 5$ LBG
sample ($Vi'z'$-LBGs); $M_\mathrm{UV}=-18.6$ is
fainter than the limiting magnitude of this sample by
1.7 mag.
\reftbl{ld} presents $\rho_L^\mathrm{obs}$ and
$\rho_L^{L>0.1L^\ast_{z=3}}$.

\begin{deluxetable}{ccccc}
  \tablecolumns{5}
  \tablewidth{0pt}
  \tablecaption{UV luminosity densities \label{tbl:ld}}
  \tablehead{
    \colhead{} & \colhead{} & \colhead{}
    & \multicolumn{2}{c}{$\rho_L$ [ergs$^{-1}$ Hz$^{-1}$ Mpc$^{-3}$]} \\
    \cline{4-5} \\
    \colhead{\hspace{0.1cm} Sample Name\hspace{0.1cm} }
    & \colhead{\hspace{0.3cm} $\bar{z}$\hspace{0.2cm} }
    & \colhead{\hspace{0.2cm} $s_z$\hspace{0.3cm} }
    & \colhead{$\rho_L^\mathrm{obs}$}
    & \colhead{$\rho_L^{L>0.1L^\ast_{z=3}}$} \\
    \colhead{} & \colhead{} & \colhead{}
    & \colhead{\hspace{0.3cm}
      ($M_\mathrm{UV}<M_\mathrm{limit}$\tablenotemark{\ast})\hspace{0.3cm} }
    & \colhead{\hspace{0.3cm} ($M_\mathrm{UV}<-18.6$)\hspace{0.3cm} }
  }
  \startdata 
$BRi'$-LBGs  & 4.0 & 0.3
& $2.21^{+0.01}_{-0.01}\times 10^{26}$ & $2.82^{+0.04}_{-0.04}\times 10^{26}$ \\
$Vi'z'$-LBGs & 4.7 & 0.3
& $4.33^{+0.03}_{-0.05}\times 10^{25}$ & $1.85^{+0.99}_{-0.58}\times 10^{26}$ \\
  \enddata 
  \tablenotetext{\ast\ }{%
Observed limiting magnitudes;
$M_\mathrm{UV}=-19.2$ for $BRi'$-LBGs and
$M_\mathrm{UV}=-20.3$ for $Vi'z'$-LBGs.}
\end{deluxetable} 

The ultraviolet luminosity density can be used to measure the
star formation rate (SFR) density in the universe.
We compute the cosmic SFR density using the relationship between
SFR and ultraviolet luminosity, $L_\mathrm{UV}$, given by
\citet{kennicutt1998}:
\begin{eqnarray}
\lefteqn{\mathrm{SFR}\ [M_\odot\ \mathrm{yr}^{-1}]}\hspace{0.1cm} \nonumber\\
&=& 1.4\times 10^{-28}\,L_\mathrm{UV}
\ [\mathrm{ergs}\ \mathrm{s}^{-1}\ \mathrm{Hz}^{-1}]. \label{eq:sfr}
\end{eqnarray}
This formula assumes a Salpeter initial mass function with
$0.1\ M_\odot<M<100\ M_\odot$.
The SFR density is corrected for dust extinction following
\citet{hopkins2004}, who used the dust extinction formula by
\citet{calzetti2000} and assumed $E(B-V)=0.128$ at all
redshifts.

\begin{figure}[t]
  \begin{center}
    \includegraphics[scale=0.4]{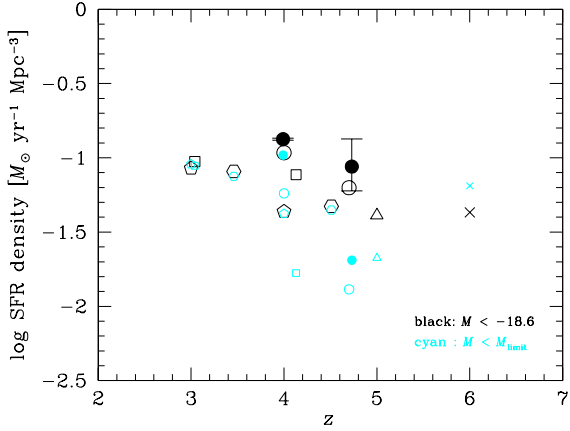}
  \end{center}
  \caption{%
Cosmic star formation rate (SFR) density as a function of redshift.
The black symbols indicate the SFR densities
calculated by integrating the luminosity functions down to
$M_\mathrm{UV}=-18.6$ (equivalent to $L=0.1L^\ast_{z=3}$
from Steidel et al. 1999),
while the cyan symbols show the lower limits,
i.e., the contribution from actually observed galaxies only.
The filled circles represent the measurements from this study;
the other points come from \citet{steidel1999} (open squares),
\citet{iwata2003} (open triangles), \citet{ouchi2004} (open circles),
\citet{gabasch2004} (open hexagons),
\citet{sawicki2006} (open pentagons),
and \citet{bouwens2005} (crosses).}
\label{fig:sfr-lbg}
\end{figure}

In \reffig{sfr-lbg}, we show the cosmic SFR densities obtained in
this study as a function of redshift (filled circles), comparing with
those calculated from LFs of LBGs at $3\le z\le 6$ in the literature
(other kinds of symbols).
We similarly integrate their LFs down to their observed limiting
luminosities to obtain $\rho_L^\mathrm{obs}$
and down to $L=0.1L^\ast_{z=3}$ to obtain
$\rho_L^{L>0.1L^\ast_{z=3}}$,
convert them to the SFR densities using
Eq.(\ref{eq:sfr}), and correct for the same amount of dust
extinction to provide a fair comparison.
The black symbols indicate $\rho_L^{L>0.1L^\ast_{z=3}}$,
while the cyan symbols are for the lower limit, $\rho_L^\mathrm{obs}$.
Note that the limiting luminosity of the LF at $z\sim 6$
by \citet{bouwens2005} is fainter than $0.1L^\ast_{z=3}$,
resulting in
$\rho_L^\mathrm{obs} > \rho_L^{L>0.1L^\ast_{z=3}}$.
It can be seen that
$\rho_L^{L>0.1L^\ast_{z=3}}$
is almost constant over
$3\lesssim z\lesssim 5$, and perhaps turns to decrease
at a certain point between $z\sim 5$ and 6.
It should be emphasized that the measurements of the SFR density
at $z\sim 4$ and $z\sim 5$ are largely improved by this study.
At $z\sim 4$, the observed SFR density is very close to
$\rho_L^{L>0.1L^\ast_{z=3}}$,
since our $BRi'$-LBG sample reaches very close to
$0.1L^\ast_{z=3}$.
Thus, there is little uncertainty in our measurement of
$\rho_L^{L>0.1L^\ast_{z=3}}$
at $z\sim 4$, while the previous measurements of SFR density
integrated to $0.1L^\ast_{z=3}$ have
uncertainties of a factor of two or more due to a large extrapolation
of the LF.
From this figure, we can robustly conclude that the cosmic SFR density
does not drop from $z\sim 3$ to $z\sim 4$, since the ``observed''
SFR density at $z\sim 4$ is not lower than
$\rho_L^{L>0.1L^\ast_{z=3}}$
at $z\sim 3$.
At $z\sim 5$, large errors owing to the uncertainty in the shape
of the LFs at faint magnitudes still remain.
However, from our observed SFR density,
we can put a stronger constraint on the ``decrease''
in the cosmic SFR density from $z\sim 4$ to $z\sim 5$;
the decrease, if any, cannot be larger than a factor of about
five.

\subsubsection{%
Luminosity Dependent Evolution of the Cosmic Star Formation Rate
Density \\
over $0\le z\le 6$}
\label{subsubsec:dis-sfr-ld}

\begin{figure}[t]
  \begin{center}
    \includegraphics[scale=0.4]{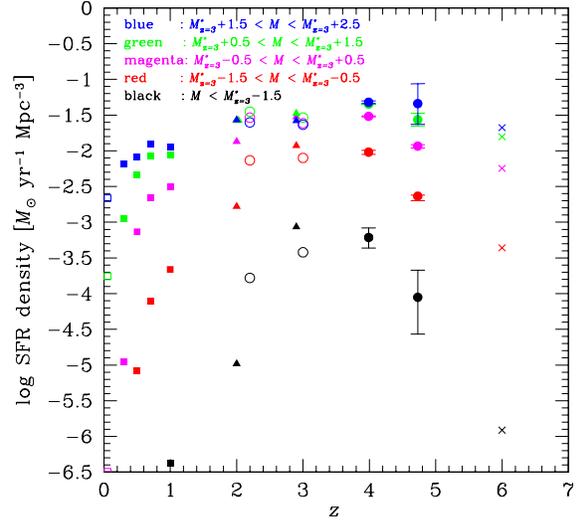}
  \end{center}
  \caption{%
Cosmic star formation rate (SFR) density as a function of redshift
over the redshift range of $0\le z\le 6$,
demonstrating its luminosity dependent evolution.
The black, red, magenta, green,and blue symbols indicate the
SFR density from galaxies with $M<M^\ast_{z=3}$,
$M^\ast_{z=3}-1.5<M<M^\ast_{z=3}-0.5$,
$M^\ast_{z=3}-0.5<M<M^\ast_{z=3}+0.5$,
$M^\ast_{z=3}+0.5<M<M^\ast_{z=3}+1.5$,
and $M^\ast_{z=3}+1.5<M<M^\ast_{z=3}+2.5$,
respectively.
The filled circles represent the measurements from this study.
The open squares comes from \citet{wyder2005},
the filled squares from Arnouts et al. (2005; GALEX),
the filled triangles from Arnouts et al. (2005; HDF),
the open circles from \citet{sawicki2006},}
and the crosses from \citet{bouwens2005}.
\label{fig:sfr-ld}
\end{figure}

We examine the evolution of the cosmic SFR densities contributed from
galaxies with different magnitudes over the redshift range of
$0\le z\le 6$.
\reffig{sfr-ld} compares their evolutionary behavior.
The black, red, magenta, green, and blue symbols indicate the
cosmic SFR densities from galaxies with magnitudes in the range
$M<M^\ast_{z=3}-1.5$,
$M^\ast_{z=3}-1.5<M<M^\ast_{z=3}-0.5$,
$M^\ast_{z=3}-0.5<M<M^\ast_{z=3}+0.5$,
$M^\ast_{z=3}+0.5<M<M^\ast_{z=3}+1.5$,
and $M^\ast_{z=3}+1.5<M<M^\ast_{z=3}+2.5$,
respectively.
The filled circles represent our measurements, and the other symbols
show the values similarly calculated using the LFs
taken from the literature.
There is a remarkable dependence of the evolutionary behavior
of the SFR density on luminosity,
as is previously noticed by \citet{shimasaku2005}
who compare the evolution of the total cosmic SFR density
and the SFR density from galaxies brighter than $M=-21.3$
to find that the SFR density from bright galaxies drastically changes
with time.
From \reffig{sfr-ld},
it is shown that the SFR density
from fainter galaxies evolves more mildly.
In addition, the peak position varies with luminosity;
the contribution from fainter galaxies peaks at earlier epochs.
For example, the SFR density for galaxies with
$M^\ast_{z=3}-1.5<M<M^\ast_{z=3}-0.5$ rises by about an order of
magnitude from $z\sim 6$ to $z\sim 4$, remains unchanged to $z\sim 3$,
and then drops by more than three orders of magnitude to the
present epoch, thus having a sharp peak at $z\sim 3$ -- 4.
On the other hand, the SFR density for galaxies with
$M^\ast_{z=3}+1.5<M<M^\ast_{z=3}+2.5$ is almost constant
within a factor of about two at $z\gtrsim 2$
with a blunt peak at $z\sim 4$ -- 5.
\citet{iwata2003}, \citet{gabasch2004}, and \citet{sawicki2006}
have reported different evolutions of the LBG
LF from $z\sim 4$ or 5 to 3 from what we find in the SDF.
At $3\lesssim z\lesssim 5$,
the luminosity density of these authors shows an opposite trend
that fainter galaxies evolve more rapidly.
However, it is at $z\lesssim 3$ and $z\gtrsim 5$ that
most drastic evolution occurs in the luminosity density.
Thus, using the LFs from these authors does not qualitatively
change the results on the overall evolution
over $0\lesssim z\lesssim 6$ obtained here.

This luminosity-dependent evolution of the cosmic SFR density
reflects the evolution of the LF discussed
in \S\ref{subsec:dis-lf}.
The strong increase in the SFR density for
galaxies brighter than $M^\ast_{z=3}-0.5$
from $z\sim 6$ to $z\sim 3$ is due to the brightening of $M^\ast$
with time.
The SFR densities for
these bright galaxies
then drop to the present epoch,
as $M^\ast$ fades during the same period.
The evolution of the SFR densities for
galaxies fainter than $M^\ast_{z=3}-0.5$
is mild,
since they are less sensitive to the change in $M^\ast$.
The earlier peak position of the SFR density of fainter galaxies
suggests that faint galaxies
are more dominant in terms of the cosmic SFR density at earlier epochs.

We cannot completely rule out the possibility that 
the luminosity-dependent evolution in the cosmic SFR density 
seen above is not real but reflects luminosity-dependent 
evolution of some other properties such as dust extinction.
If, for instance, galaxies have the smallest dust extinction 
at $z\sim 3$ -- 4, then the observed $M^\star$ will be 
the brightest at these redshifts, as seen in \reffig{schechter}, 
even if the intrinsic, dust-corrected $M^\star$ does not 
change with redshift.
In this case, the behavior of the luminosity-dependent SFR 
density will be qualitatively similar to that found in \reffig{sfr-ld}.
However, no observation has reported strong evolution of $E(B-V)$ 
at least for LBGs.
The $E(B-V)$ distribution of LBGs appears to be unchanged 
over $z\sim 5$ and 3 \citep{iwata2003,ouchi2004}.
In addition, $E(B-V)$ is known to be independent 
of apparent $M$ \citep{adelberger2000,ouchi2004}.

\subsection{Specific Star Formation Rate} \label{subsec:dis-ssfr}

We explore the evolution of
what we define as the specific SFR here, i.e.,
the cosmic SFR per unit baryon mass in dark haloes.
The specific SFR means the efficiency of star formation averaged over
the all dark haloes present at a given redshift.
The specific SFR is computed by dividing the cosmic SFR density,
$\rho_L^{L>0.1L^\ast_{z=3}}$,
by the average mass density of baryons
confined in dark haloes, $\rho_\mathrm{b}$.
Here, $\rho_\mathrm{b}$ is calculated by
\begin{eqnarray}
\rho_\mathrm{b}(z) &=&
\frac{\mathrm{\Omega}_b}{\mathrm{\Omega}_m}
\int^\infty_{M_\mathrm{min}(z)}n(M, z)M\,dM,
\end{eqnarray}
where $\mathrm{\Omega}_b=0.05$ is the density parameter of baryons,
$n(M, z)$ is the mass function of dark haloes at redshift $z$ predicted
by the standard Cold Dark Matter model with $\sigma_8=0.9$,
$M_\mathrm{min}(z)$
is the halo mass corresponding to a virial temperature of
$1\times 10^4$ K.
We assume here that gas does not cool (and thus stars are not formed)
in dark haloes with $T< 1\times 10^4$ K.
Note that $M_\mathrm{min}(z)$ is much smaller than the estimated
mass of haloes which host the faintest galaxies we account here,
i.e., galaxies with $0.1L^\ast_\mathrm{z=3}$.

\begin{figure}[t]
  \begin{center}
    \includegraphics[scale=0.4]{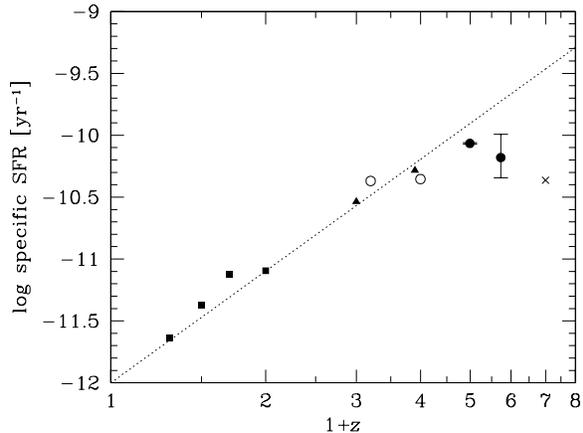}
  \end{center}
  \caption{%
Specific star formation rate (SFR) versus $(1+z)$ over $0\le z\le 6$.
Symbols are the same as in \reffig{sfr-ld}.
The dotted line shows evolution in power-law,
specific $\mathrm{SFR}\propto (1+z)^3$.
\label{fig:ssfr}}
\end{figure}

\reffig{ssfr} shows the evolution of the specific SFR over
$0\le z\le 6$.
The filled circles represent our measurements
and the other symbols show the values from the literature.
The dotted line in the figure shows $(1+z)^3$ evolution.
We find that the specific SFR
increases in proportion to $(1+z)^3$ up to $z\sim 4$.
In other words, the star formation efficiency in the universe
rises with redshift.
At $z\gtrsim 4$, on the other hand, the specific SFR appears
to decrease.

Here, according to the standard Cold Dark Matter model,
the average internal density in physical units
of baryons in dark haloes virialized at
a given redshift,
$\left<\rho_\mathrm{b}^\mathrm{DH}\right>$,
is known to be approximately proportional
to the average mass density of the universe at that time,
and the average mass density of the universe changes as $(1+z)^3$.
Therefore, it implies that the specific SFR increases approximately
in proportion to $\left<\rho_\mathrm{b}^\mathrm{DH}\right>$
up to $z\sim 4$.
To put it another way,
since the specific SFR is the internal density of SFR in dark haloes,
$\left<\rho_\mathrm{SFR}^\mathrm{DH}\right>$,
divided by
$\left<\rho_\mathrm{b}^\mathrm{DH}\right>$,
this result can be expressed
in terms of $\left<\rho_\mathrm{SFR}^\mathrm{DH}\right>$
and $\left<\rho_\mathrm{b}^\mathrm{DH}\right>$ as:
\begin{eqnarray}
\left<\rho_\mathrm{SFR}^\mathrm{DH}\right>
&\propto& \left<\rho_\mathrm{b}^\mathrm{DH}\right>^2.
\label{eq:schmidt}
\end{eqnarray}
It is surprising that the star formation in galaxies obeys
such a simple power law over $90\%$ of the cosmic history
($0<z\lesssim 4$).
Equation (\ref{eq:schmidt}) resembles the Schmidt law
\citep{schmidt1959} of
nearby disk galaxies, a relationship between the disk-averaged
SFR and cold gas surface densities of
$\Sigma_\mathrm{SFR} = \mathrm{const.}\times\Sigma_\mathrm{gas}^N$
with $N \sim 1.4$ (Kennicutt 1998).

We present here one possible 
explanation of the observed behavior of the specific SFR.
The analytical modeling of galaxy formation based on the CDM 
model by \citet{hernquist2003} shows that the gas 
cooling rate, $\dot{M}_\mathrm{cool} / M_\mathrm{DH}$, of dark haloes 
roughly scales with $H(z)^2$, or equivalently with $(1+z)^3$ 
as $H(z)$ is approximately proportional to $(1+z)^{3/2}$ in 
our cosmology. 
They also find that the star formation rate in dark haloes 
scales with the cooling rate at low and intermediate redshifts, 
while it becomes to be no longer dependent on the cooling rate 
at redshifts larger than a certain limit, where the internal 
density of haloes is so high that the cooling time is shorter 
than the consumption rate of cooled gas into stars.
If we adopt this modeling, then our result implies that 
the cosmic SFR is primarily determined by the gas cooling 
rate at $z \lesssim 4$, while it is governed by the conversion 
rate of cooled gas into stars at larger redshifts.

Of course, the cooling rate may not be the major factor 
determining the cosmic star formation rate.
There are many other factors on which the star formation 
in galaxies can largely depends, 
such as the feedback by galactic winds, the UV background 
radiation field, and environmental effects like galaxy 
merging and interactions.
Since the relative importance of these factors will 
vary with the mass of dark haloes, 
it will be interesting to measure the specific SFR 
as a function of halo mass.

\section{SUMMARY AND CONCLUSIONS} \label{sec:sum}

We investigated the luminosity functions and star formation rates
of Lyman-break galaxies (LBGs) at
$z\sim 4$ and 5 based on the optical imaging data obtained in the
Subaru Deep Field (SDF) Project.
The SDF Project is a program conducted by Subaru Observatory
to carry out a deep and wide galaxy survey in the SDF.

Three samples of LBGs in a contiguous 875 arcmin$^2$ area
are constructed.
One consists of LBGs at $z\sim 4$ selected with the \br\ vs \ri\ 
diagram ($BRi'$-LBGs).
The other two consist of LBGs at $z\sim 5$ selected with two kinds of
two-color diagrams: \vi\ vs \iz\ ($Vi'z'$-LBGs) and \ri\ vs
\iz\ ($Ri'z'$-LBGs).
The number of LBGs detected is $3,808$ for $BRi'$-LBGs, 539 for
$Vi'z'$-LBGs, and 240 for $Ri'z'$-LBGs.
The adopted selection criteria are proved to be fairly reliable
by the spectroscopic data.
We use Monte Carlo simulations to estimate the completeness and
the fraction of contamination by interlopers.
The redshift distribution functions obtained by the simulations
are well consistent with the redshift distributions of
spectroscopically identified objects in the LBG samples.

We derived the luminosity functions at rest-frame ultraviolet
wavelengths down to as faint as $M_\mathrm{UV}=-19.2$ at $z\sim 4$ and
$M_\mathrm{UV}=-20.3$ at $z\sim 5$.
The cosmic star formation rate (SFR) density was measured from the
ultraviolet luminosity density derived by integrating the luminosity
functions.
On the basis of the observed SFR density
and the standard Cold Dark Matter model, the cosmic SFR
per unit baryon mass in dark haloes, i.e., the specific SFR, is
computed.
Combining the results obtained in this work with those taken from the
literature, we reached the following conclusions:

\renewcommand{\labelenumi}{(\roman{enumi})}
\begin{enumerate}

\item
Clear evolution of the luminosity function is detected.
The characteristic magnitude $M^\ast$ changes strongly
over the redshift range of $0\le z\le 6$.
It brightens by about 1 mag from $z\sim 6$ to $z\sim 4$,
remains unchanged to $z\sim 3$ and then
fades by about 3 mag
from $z\sim 3$ to $z\sim 0$.
On the other hand, the normalization factor, $\phi^\ast$,
and the faint-end slope, $\alpha$, are almost constant
over $0.5\lesssim z\lesssim 6$.

\item
The measurements of the SFR density at $z\sim 4$ and $z\sim 5$ are
greatly improved, since the luminosity functions are derived down to
very faint magnitudes.
Hence, the evolutionary behavior of the SFR density is more
constrained.
The cosmic SFR density does not drop from $z\sim 3$ to $z\sim 4$,
and the decrease from $z\sim 4$ to $z\sim 5$, if any, cannot be
larger than a factor of about five.

\item
The SFR density contributed from brighter
galaxies changes more drastically with cosmic time.
The contribution from
galaxies brighter than $M^\ast_{z=3}-0.5$
has a sharp peak around
$z=3$ -- 4,
while that from
galaxies fainter than $M^\ast_{z=3}-0.5$
evolves relatively mildly
with a broad peak at earlier epoch.
This luminosity-dependent evolution of the cosmic SFR density suggests
that the galaxy population that contributes to the total SFR
density in the universe varies with time, reflecting the
evolution of the luminosity function.

\item
The specific SFR is found to scale with redshift as $(1+z)^3$
up to $z\sim 4$, implying that
the efficiency of star formation is on average
higher at higher redshift in proportion to the cooling rate
within dark haloes.
It seems that this is not simply the case at $z\gtrsim 4$.
\end{enumerate}

\acknowledgments
We would like to thank the referee for valuable comments and
suggestions.
We are deeply grateful to the Subaru Telescope staff for their
devoted support in this project.


\end{document}